\documentclass[12pt]{article}
\usepackage{graphicx}
\usepackage{geometry}
\usepackage{amssymb}
\usepackage{setspace}
\usepackage[dvipsnames]{pstricks}
\usepackage{pst-node}
\usepackage{pst-slpe}
\usepackage{pst-tree}
\usepackage{bbm}
\usepackage{amsmath, amsthm, amssymb}
\usepackage{pgfplots}
    \pgfplotsset{width=8cm,compat=1.5.1}
\usetikzlibrary{arrows.meta}
    \usetikzlibrary{shapes,snakes}
\usepackage{amsmath,amssymb,amsthm}
\usepackage{tikz}
\usepackage{caption}
\usepackage{subcaption}
\usepackage{booktabs}
\usepackage{empheq}
\usepackage[most]{tcolorbox}
\usepackage{diagbox}
\usepackage{blkarray}

\addtocontents{toc}{\protect\setcounter{tocdepth}{1}}
\appto\appendix{\addtocontents{toc}{\protect\setcounter{tocdepth}{1}}}

\newtcbox{\mymath}[1][]{%
    nobeforeafter, math upper, tcbox raise base,
    enhanced, colframe=blue,
    colback=yellow!30, boxrule=1pt,
    #1}
    
\usepackage{natbib}
\usepackage{draftwatermark}
\SetWatermarkText{}
\SetWatermarkScale{4}

\newtheorem{theorem}{Theorem}
\newtheorem{algorithm}{Algorithm}
\newtheorem{axiom}{Axiom}

\newtheorem{example}{Example}

\newtheorem{lemma}{Lemma}

\newtheorem{definition}{Definition}
\usepackage{tikz}
\renewcommand{\emptyset}{\varnothing}

\title{Consumption Dependent Random Utility
\footnote{I am grateful to Victor Aguiar, Roy Allen, Miguel Ballester, Laurent Bouton, Simone Cerreia-Vioglio, Edward Honda, Sean Horan, Roger Lagunoff, Jay Lu, Andrew Mackenzie, Kyle Monk, Tianshi Mu, Collin B. Raymond, John Rehbeck, Marciano Siniscalchi, and Joshua Teitelbaum  as well as seminar participants at Bocconi, BRIC 2023, Bristol, Georgetown, NASMES 2023, Queensland, RUD 2024, and SAET 2023 for their helpful comments during the course of this project. I am especially grateful to Peter Caradonna, Christopher Chambers, and Yusufcan Masatlioglu for their continued support and insightful conversations throughout the course of this project. Much of the content of this paper was circulated previously in my job market paper ``Random Utility, Repeated Choice, and Consumption Dependence". \\
Turansick: Department of Decision Sciences and IGIER, Universit\'{a} Bocconi.  E-mail:  \texttt{christopher.turansick@unibocconi.it}}}
\author{Christopher Turansick}
\date{\today}

\begin{document}

\maketitle

\begin{abstract}
    We study a dynamic random utility model that allows for consumption dependence. We axiomatically analyze this model and find insights that allow us to distinguish between behavior that arises due to consumption dependence and behavior that arises due to state dependence. As part of our analysis, we show that it is impossible to distinguish between myopic and dynamically sophisticated agents when there are well defined marginal choices in each period. Building on our axiomatic analysis, we develop a revealed preference test for consumption dependent random utility. Our test can be implemented with real data, and we show that our test offers computational improvements over the natural extension of \citet{kitamura2018nonparametric} to our environment.
\end{abstract}

\onehalfspacing
\section{Introduction}\label{Intro}
People are heterogeneous in their choices. As a result, population level choice data often appears stochastic to an analyst. In economics, a common model of stochastic choice data is the random utility model. It supposes there is some distribution over preferences which induces the observed distribution over choices. This distribution over preferences captures heterogeneity in a population. Each agent has a type which induces a deterministic preference which is known to them. The fact that an agent's type is not known to the analyst induces the apparent randomness of choice. Since the work of \citet{rust1987}, there has been much interest in using dynamic versions of the random utility model in empirical applications.

In this paper, we study a dynamic extension of the random utility model which allows for consumption and state dependence. Consumption dependence is the idea that an agent's preference today may depend on their choice history. State dependence is the idea that an agent's preference may depend on some exogenous underlying state of the world which varies over time. While we allow for both consumption and state dependence, our focus is on consumption dependence. Our analysis focuses on characterizing which dynamic choice datasets are consistent with a nonparametric model of dynamic random utility which allows for both of these dynamic forces. We begin with an axiomatic approach which sheds light on which intertemporal behaviors are ruled out by our consumption dependent random utility model. We complement the axiomatic approach with a positive approach which builds to an implementable test of our model.

Our motivating examples of consumption dependence are habit formation and learning through experience. Habit formation captures the idea that taking an action or consuming a good today makes it easier to take that same action tomorrow. Habit formation has been studied in many different economic contexts including rational addiction \citep{becker1988theory}, consumer choice \citep{pollak1970habit}, and growth and savings models \citep{carroll2000saving}. Going back to \citet{yerkes1908relation}, there has been much evidence in the psychology literature on the presence of habit formation in human choice. \citet{havranek2017habit} offers a recent review of the evidence of habit formation in economic contexts. Learning through experience captures the following phenomenon. Suppose a consumer has beliefs about their consumption utility of a good. These beliefs influence a consumer's initial choice. Upon consumption of the good, the consumer faces some utility realization and are able to update their beliefs. The key point here is that learning is done only through consumption of the good. Learning through experience has seen much study in the marketing literature \citep{hoch1986consumer,hoch1989managing}.

In the main text, we consider a two period model of choice which we call the consumption dependent random utility model (CDRUM). In the first period there is some distribution over preferences which represents a population of heterogeneous but individually rational agents. Each agent chooses from a menu of alternatives in order to maximize their preference. In the second period, each agent's preference is realized according to a transition function. A transition function $t$ takes as input an agent's choice and preference today and returns a distribution over preferences tomorrow. As tomorrow's preference depends, in a potentially heterogeneous manner, on today's choice, these transition functions are how we capture consumption dependence in our model. Upon realizing their second period preference, agents once again choose from a menu of alternatives in order to maximize their preference. We study this model when an analyst observes a random joint choice rule. A random joint choice rule records how frequently $x$ is chosen in period one and $y$ is chosen in period two conditional on the realized menus $A$ and $B$ in the first two respective periods.

Our first set of results give the identification properties of CDRUM and axiomatically characterizes which behaviors are consistent with CDRUM. CDRUM is characterized by two axioms. The first axiom we call complete monotonicity. This axiom is an extension of the classic static random utility axiom of \citet{block1959random} and \citet{falmagne1978representation} extended to a dynamic environment. Intuitively, it is a statement about gross substitutes written in terms of choice frequencies. It captures the fact that the choice frequency of $(x,y)$ should never increase if another alternative is made available. Here $x$ corresponds to first period choice and $y$ corresponds to second period choice. Complete monotonicity extends this intuition to take into account the partial order structure of set inclusion. The second axiom is called marginality. Marginality asks that first period choice frequencies are independent of second period menus. Note that marginality still allows for arbitrary history dependence. History dependence is the statistical dependence between first and second period choice. As we are modeling consumption dependent behavior, we need to allow for perfect correlation between first and second period choice. However, marginality does not allow for an agent in the first period to condition their choice on the realization of their second period menu.

Our second result compares the behavioral content of myopia with that of dynamic sophistication. We consider an extension of CDRUM in which an agent realizes, in the first period, a preference over their two period consumption stream. This model allows for dynamic sophistication while CDRUM inherently focuses on myopic agents. Once we allow for dynamic sophistication, behavior can violate both complete monotonicity and marginality. However, if we impose marginality, then dynamically sophisticated behavior becomes observationally equivalent to CDRUM. We use this to motivate our usage of random joint choice rules over the more typical conditional choice probabilities. Our result tells us that, when a random joint choice rule induces a well defined system of conditional choice probabilities, myopia and dynamic sophistication are observationally equivalent.

Our next result axiomatizes a refinement of CDRUM. Specifically, we consider CDRUM without any state dependence, thus capturing pure (heterogeneous) consumption dependence. In this case, our model is characterized by one additional axiom. We call this axiom choice set independence. Choice set independence asks that, whenever we can define second period choice probabilities conditional on first period choice and menu, second period conditional choice frequencies are independent of the agent's first period menu. This axiom captures the fact that when an agent is purely choice dependent, the only component of the first period that matters for second period choice is the agent's first period choice.

Our final set of results take our axiomatic characterization of CDRUM and translates it into a revealed preference style test of CDRUM, potentially allowing for some menus to be unobserved. We develop two tests which extend modern tests of the static random utility model. Our first test extends the techniques of \citet{kitamura2018nonparametric} while our second test extends the techniques of \citet{turansick2023alternative}. The intuition behind these hypothesis tests comes from the fact that a convex polytope can be represented as convex combination of its extreme points, the basis for the test of \citet{kitamura2018nonparametric}, or it can alternatively be represented by the intersection of finitely many half-spaces, the basis for the test of \citet{turansick2023alternative}. It turns out that axioms we use to characterize CDRUM correspond to the half-space defining inequalities of CDRUM when we observe every menu of alternatives. When some menus are unobserved, we utilize a change of variables and slack variables in order to provide an existential linear program characterizing CDRUM. The existence problem for this linear program can be directly tested using the statistical techniques developed in \citet{fang2023inference}. While the test based on \citet{turansick2023alternative} offers computational improvements over the test based on \citet{kitamura2018nonparametric}, when utilizing the column generation procedure of \citet{smeulders2021nonparametric}, the min max computational burden of the extreme point test is of the same order as the computational burden of the half-space test.

The rest of this paper is structured as follows. In Section \ref{Model} we formally introduce our notation and model. In Section \ref{sec:axioms} we axiomatically characterize our model. In Section \ref{sec:hypo} we develop two tests of our model. Finally, we conclude in Section \ref{sec:conc} and discuss the related literature. All proofs are relegated to the appendix.

\section{Model}\label{Model}
\subsection{Primitives}
Let $X$ be a finite set of alternatives with typical elements $x, y,$ and $z$. We use $\mathcal{X}$ to denote the collection of nonempty subsets of $X$. $\mathcal{L}(X)$ denotes the set of linear orders of $X$ with typical element $\succ$. We let $\Delta(\mathcal{L}(X))$ denote the set of probability distributions over $\mathcal{L}(X)$ with typical element $\nu$. We use $M(\succ,A)$ to denote the element $x \in A$ that maximizes $\succ$ in $A$. Further, we use the shorthand to $x \succ A$ to denote that $x \succ y$ for all $y \in A$ with $A \succ x$ defined analogously. Define $N(x,A)=\{\succ|x \succ A \setminus \{x\}\}$ and $I(x,A)=\{\succ|X \setminus A \succ x \succ A \setminus\{x\}\}$. $N(x,A)$ denotes the set of linear orders maximized by $x$ in $A$ and $I(x,A)$ denotes the set of linear orders maximized by $x$ in $A$ but not maximized by $x$ for all $A \cup \{z\}$ with $z \not \in A$.

\subsection{Data Generating Process}
For expositional reasons, in the main text, we consider a two period model of choice. The model and results in Section \ref{sec:axioms} are extended to an arbitrary finite number of periods in the appendix. The aim of our model is to capture two forces in dynamic choice. The first force is consumption dependence which asks that today's preference depends on yesterday's choice. The second force is state dependence which asks that today's preference depends on today's state of the world. We capture both of these forces through what we call a transition function.

\begin{definition}
    We call a function $t:X \times \mathcal{L}(X) \rightarrow \Delta(\mathcal{L}(X))$ a \textbf{transition function}.
\end{definition}

We use the notation $t_{\succ'}(x,\succ)$ to correspond to the probability weight put on $\succ'$ when $x$ and $\succ$ are the inputs of $t$. The role of a transition function is to take in an agent's chosen alternative and preference today and return a distribution over preferences tomorrow. The choice input in a transition function captures consumption dependence. The preference input acts a reduced form method of capturing the state of the world today. In the scope of our model, a transition function is the (stochastic) mapping between an agent's preference and choice in the first period and their preference in the second period. Our model of choice proceeds in the following manner. In the first of two periods, each agent in a population of agents chooses the alternative which maximizes their preference $\succ$ which is realized from distribution $\nu$. Given their choice and preference in the first period, each agent realizes a preference according to a transition function $t$. The agent then chooses the alternative which maximizes their preference in the second period. Our goal is to characterize which datasets are consistent with this data generating procedure.

\begin{definition}
A function $p:X^2 \times \mathcal{X}^2 \rightarrow [0,1]$ is a \textbf{random joint choice rule} (rjcr) if it satisfies the following.
\begin{enumerate}
    \item $p(x,y,A,B) \geq 0$
    \item $\sum_{x \in A}\sum_{y \in B}p(x,y,A,B)=1$
\end{enumerate}
\end{definition}

A random joint choice rule corresponds to the dataset observable to an analyst. The rjcr $p(x,y,A,B)$ captures the frequency with which alternative $x$ is chosen from menu $A$ in the first period and alternative $y$ is chosen from menu $B$ in the second period. These frequencies can be thought of as arising from a population of agents making decisions across two time periods. We note that random joint choice rules are a stronger type of data than what is sometimes used in dynamic discrete choice settings. In dynamic discrete choice, commonly an analyst observes a system of conditional choice probabilities. Formally, a system of conditional choice probabilities consists of a random choice rule $p(x,A)$ over first period choices and conditional random choice rules $p(y,B|x,A)$ over second period choices. Every system of conditional choice probabilities can be represented as a random joint choice rule, but there are random joint choice rules which cannot be represented by conditional choice probabilities. One of our goals in this paper is to offer an axiomatic characterization of a consumption dependent random utility model. While we present each of our axioms in terms of an underlying rjcr, each of our axioms can be applied to test conditional choice probabilities by simply transforming the conditional choice probabilities into their rjcr representation.

\begin{definition}\label{CDRUMdef}
    A random joint choice rule $p$ is \textbf{consistent} with the consumption dependent random utility model (CDRUM) if there exists a probability distribution over preference $\nu$ and a transition function $t$ such that the following holds for all $A,B \subseteq X$ and $(x,y) \in A\times B$.
    \begin{equation}\label{CDRUMEq}
        p(x,y,A,B)=\sum_{\succ \in N(x,A)} \sum_{\succ' \in N(y,B)}\nu(\succ)t_{\succ'}(x,\succ)
    \end{equation}
\end{definition}

Definition \ref{CDRUMdef} is the formal definition of the model we described earlier. For a rjcr $p$ to be consistent with CDRUM, we ask that there exists some distribution over preferences governing first period choice and there exists some transition function which governs second period choice. We assume rationality of our agents in that they maximize their realized preference in each period.

In addition to CDRUM, which captures both consumption and state dependence, we also wish to consider a model of pure consumption dependence. In other words, we also consider a specification of CDRUM without state dependence.

\begin{definition}\label{SIT}
    We call a transition function $t:X \times \mathcal{L}(X) \rightarrow \Delta(\mathcal{L}(X))$ \textbf{state independent} if, for all $x \in X$, $t(x,\succ)=t(x,\succ')$ for all $\succ, \succ' \in \mathcal{L}(X)$. When $t$ is state independent, we will often write $t(x)$ instead of $t(x,\succ)$.
\end{definition}

State independent transition functions exactly capture pure consumption dependence. In order to distinguish between the effects of state and consumption dependence, we are also interested in characterizing datasets which arise due purely to consumption dependence. In terms of our model, these are datasets which consistent with CDRUM with a state independent transition function. 

\begin{definition}\label{SICDRUM}
    We say that a rjcr $p$ is \textbf{consistent} with the state independent consumption dependent random utility model (SI-CDRUM) if there exists a state independent transition function $t$ such that following holds for all $A,B \subseteq X$ and $(x,y) \in A\times B$.
    \begin{equation}\label{SICDRUMEq}
        p(x,y,A,B)=\sum_{\succ \in N(x,A)} \sum_{\succ' \in N(y,B)}\nu(\succ)t_{\succ'}(x)
    \end{equation}
\end{definition}

Before moving on, we note that our model does not capture an important force in dynamic choice. Our model does not capture any form of planning or dynamic sophistication by the agents. In the first period of choice, the distribution over preference $\nu$ does not depend on the menu that the agent faces in the second period. This means that our agents are best interpretted as myopic and unable to condition on the future.

\section{Characterization and Axiomatics}\label{sec:axioms}
In this section we provide axioms which characterize CDRUM and SI-CDRUM. CDRUM is characterized by two axioms. The first is an extension of the classic random utility axiom extended to multiple time periods. The second axiom is a direct result of the myopic nature of our agents. SI-CDRUM is characterized by one further axiom which directly captures second period choices being purely a result of first period choices.

\subsection{Consumption and State Dependence}

We now study the consumption dependent random utility model. As mentioned prior, one of the characterizing axioms of CDRUM is an extension of the classic random utility axiom to multiple time periods. Just as is the case with static random utility, the concept of a M\"{o}bius inverse is important for the study of CDRUM. Given a function $f:\mathcal{X} \rightarrow \mathbb{R}$, the M\"{o}bius inverse of $f$ is recursively given by $f(A)=\sum_{A \subseteq B}g(B)$. The M\"{o}bius inverse $g(B)$ captures exactly how much is being added to or subtracted from $f(\cdot)$ at the set $B$. We consider the M\"{o}bius inverse of $p(x,y,\cdot,\cdot)$ in $\mathcal{X}^2$.\footnote{The closed form expression for the M\"{o}bius inverse here is given by $q(x,y,A,B)=\sum_{A \subseteq A'} \sum_{B \subseteq B'}(-1)^{|A' \setminus A|+|B'\setminus B|}p(x,y,A',B')$.}

\begin{equation}\label{MobInverse}
    p(x,y,A,B)= \sum_{A \subseteq A'} \sum_{B \subseteq B'}q(x,y,A',B')
\end{equation}

In classic random utility, there is a connection between $q(x,A)$, the M\"{o}bius inverse of $p(x,A)$, and the underlying distribution over preferences. Specifically, recalling that $I(x,A)=\{\succ|X \setminus A \succ x \succ A \setminus \{x\}\}$, for rationalizable data, $q(x,A)=\nu(I(x,A))$. There is a similar connection between $q(x,y,A,B)$ and our representation in CDRUM.

\begin{theorem}\label{Uniqueness}
    A distribution over preferences $\nu$ and a transition function $t$ are a consumption dependent random utility representation for random joint choice rule $p$ if and only if, for every $A,B \subseteq X$ and for all $(x,y)\in A \times B$, we have the following.
    \begin{equation}\label{CDRUMUniqueness}
        q(x,y,A,B)=\sum_{\succ \in I(x,A)}\sum_{\succ' \in I(y,B)}\nu(\succ)t_{\succ'}(x,\succ)
    \end{equation}
\end{theorem}

Theorem \ref{Uniqueness} is our identification theorem. The introduction of transition functions further complicates the standard random utility identification problem. However, the interpretation of the identification result is similar in our model. The M\"{o}bius inverse $q(x,y,A,B)$ captures the probability weight put on preferences in $I(x,A)$ and $I(y,B)$. Given that, for data consistent with CDRUM, $q(x,y,A,B)$ is equal to a probability weight, it is easy to see that $q(x,y,A,B)$ must be non-negative. In fact, the standard random utility model is characterized by $q(x,A) \geq 0$.\footnote{Random utility was first characterized by \citet{falmagne1978representation}. At the time, Falmagne called the M\"{o}bius inverse function the Block-Marschak polynomials as the non-recursive form of these functions were first introduced in \citet{block1959random}. It wasn't realized until much later that the Block-Marschak polynomials are in fact just the M\"{o}bius inverse of choice probabilities.} We need more than this to characterize CDRUM, but our first axiom does ask that the M\"{o}bius inverse $q$ is non-negative.\footnote{Formally, non-negativity of a M\"{o}bius inverse is related to the concept of complete monotonicity in harmonic analysis. The set of completely monotone functions is convex and its extreme points are exactly the step functions. The extreme points of static random utility correspond to maximization of linear orders. The choice functions induced by linear orders are characterized by the step functions. The extreme points of consumption dependent random utility are contained by the set of step functions but do not include every step function. As such, we need further restrictions beyond complete monotonicity. See \citet{berg1984harmonic} for a reference on harmonic analysis.}

\begin{axiom}[Complete Monotonicity]\label{compmon}
    A random joint choice rule $p$ satisfies \textbf{complete monotonicity} if, for every $A,B \subseteq X$ and every $(x,y)\in A \times B$, we have $q(x,y,A,B) \geq 0$.
\end{axiom}

As mentioned earlier, the M\"{o}bius inverse $g(A)$ of a function $f$ captures how much is being added to or removed from $f$ at set $A$. With this in mind, complete monotonicity is simply asking that each choice set $A \times B$ is adding a non-negative amount to the choice probability of $(x,y)$. Interpreting this in the context of a population of agents, complete monotonicity says that our data is consistent with the story that every agent who chooses $(x,y)$ from a superset of $A \times B$ still chooses $(x,y)$ at $A \times B$.\footnote{A problem present in stochastic choice that is not present in deterministic choice is the fact that we as analysts are unable to connect one agent's choice in menu $A$ to their choice in menu $B$. We simply observe a distribution over choices in these menus. As such, the axiomatic exercise is unable to shed light on any one agent's behavior, but rather informs us about population level behavior. With this in mind, we interpret our complete monotonicity axiom in an as if sense. That is to say, the population level behavior is consistent with our story about each individual's behavior} As is the case with standard random utility, complete monotonicity implies regularity, $p(x,y,A,B) \geq p(x,y,A',B')$ when $A\times B \subseteq A' \times B'$. Now that we are considering choice over multiple periods, complete monotonicity has new implications not present in the static problem. Notably, complete monotonicity now implies that choice probabilities satisfy an across period increasing differences condition.
\begin{equation}\label{IncDiff}
    A\times B \subseteq A' \times B' \implies p(x,y,A,B)-p(x,y,A,B') \geq p(x,y,A',B) -p(x,y,A',B')
\end{equation}
In this instance, we find it instructive to show what type of behavior increasing differences, and thus complete monotonicity, precludes.
\begin{example}[Temptation and Waning Self Control]\label{temptation}
    Consider a population of agents who face temptation and potentially resist that temptation with self control, as is studied in \citet{gul2001temptation}. Now suppose that these agents are subject to waning self control. That is to say, as each agent repeatedly faces temptation, it becomes more difficult to exert self control. Our agents are faced with two potential choice sets; $A$ and $A \cup \{c\}$. In this case, our agents are tempted to have cake, denoted with $c$, instead of having a healthier option in $A$. When first faced with cake, these agents can perfectly resist temptation. However, if they are faced with temptation a second time, they are completely unable to resist temptation. For $x \in A$, as agents resist temptation the first time they face temptation, we get the following.
    \begin{equation*}
        p(x,x,A,A)-p(x,x,A,A \cup \{c\}) = 0
    \end{equation*}
    On the other hand, since our agents always fail to resist temptation the second time they face it, we get the following when $x$ is chosen with positive probability.
    \begin{equation*}
        p(x,x,A\cup\{c\},A)-p(x,x,A\cup\{c\},A\cup\{c\}) > 0
    \end{equation*}
    This is a failure of our increasing differences condition and thus waning self control is not consistent with complete monotonicity.
\end{example}

More generally, complete monotonicity prevents the ``mere-exposure" effect from being observed in the data (see \citet{bornstein1992stimulus} and \citet{zajonc2001mere}). The mere-exposure effect is the idea that the presence of an irrelevant alternative in an agent's menu in one period impacts the agent's choice in a following period. In the context of Example \ref{temptation}, the presence of cake in the first period makes our agents unable to resist choosing cake in the second period. As \citet{frick2019dynamic} note, the mere-exposure effect can be thought of as a dynamic analogue to many similar static effects. These effects include the attraction effect and the decoy effect which correspond to the choice frequency of one alternative increasing/decreasing in response to the addition of an irrelevant alternative. Just as regularity and the static analogue of complete monotonicity rule out the attraction and decoy effect, our dynamic analogue of complete monotonicity rules out the mere-exposure effect.\footnote{\citet{frick2019dynamic} also rule out the mere-exposure effect in their random dynamic expected utility model. However, the axiom they use to rule out the mere-exposure effect does not have an obvious relation to static complete monotonicity.}

Our next axiom puts restrictions on the dependence between first and second period choice. Recall that our agents can be thought of as myopic. This means that when an agent makes a decision in the first period, they do not take into account their choice or choice set in the second period. As such, it should be the case that we can define first period choice probabilities independently of the second period's choice set. Marginality asks exactly that.\footnote{A stronger form of marginality was introduced and studied in \citet{strzalecki2024stochastic}, \cite{li2021axiomatization}, and \citet{chambers2024correlated}. The stronger version asks not only that first period choices can be defined independently of the second period choice set but also that second period choices can be defined independently of the first period's choice set.}

\begin{axiom}[Marginality]\label{marginality}
    A random joint choice rule $p$ satisfies \textbf{marginality} if, for every $A,B,C \subseteq X$ and for every $x \in A$, we have the following.
    \begin{equation}\label{margEq}
        \sum_{y \in B}p(x,y,A,B) =\sum_{y \in C}p(x,y,A,C)
    \end{equation}
\end{axiom}

Even though marginality restricts choices so that first period choices are independent of the second period's choice set, marginality still allows for very strong forms of correlation across actual choice.

\begin{example}[Perfect Correlation and Habit Formation]\label{CorrelationConsDep}
    Consider a population of agents who are subject to strong habit formation. If an agent chooses $x$ today, then their most preferred alternative tomorrow is $x$ with certainty. This can be modeled through the following restriction on the underlying transition function.
    \begin{equation*}
        t_{\succ}(x)=\begin{cases}
            >0 \text{ if } x \succ y \text{ } \forall y \neq x \\
            0 \text{ otherwise}
        \end{cases}
    \end{equation*}
    Now suppose that in the first period, our agents prefer $x$ to $y$ half of the time. This leads to the following choice probabilities when our agents are faced with $\{x,y\}$ in the first period.\\
    \begin{table}[h!]
    \centering
    \begin{tabular}{cc}
\toprule
{$ \begin{array}{c|ccc}
 & x & y  \\
\cline{1-3} 
x & 0.5 &  0\\
y & 0 &  0.5\\
\end{array}$} & {$\begin{array}{c|cc}
 & x  \\
\cline{1-2} 
x & 0.5 \\
y & 0.5 \\
\end{array}$}\\
\bottomrule
\end{tabular}
\end{table}\\

Above, each row represents choice of an alternative in the first period and each column represents choice of an alternative in the second period. Notably, the agents' choices are perfectly correlated when they face $\{x,y\}$ in both the first and second period. Further, when the agents face $\{x\}$ in the second period, they still choose $y$ half of the time. If, instead, we consider agents who have a strong preference for variety, we can achieve perfect negative correlation between choice.
\end{example}

\begin{example}[Perfect Correlation through State Dependence]\label{CorrelationStateDep}
    Consider a population of agents, half of whom live in a city where it always rains and half of whom live in a city where it never rains. The agents who live where it rains prefer to wear rain jackets. The agents who live where it never rains prefer to wear tee-shirts. This can be modeled through the following restriction on the underlying transition function.
        \begin{equation*}
        t_{\succ'}(x,\succ)=\begin{cases}
            1 \text{ if } \succ = \succ'\\
            0 \text{ otherwise}
        \end{cases}
    \end{equation*}
    This setup induces the following choice probabilities.
        \begin{table}[h!]
    \centering
    \begin{tabular}{cc}
\toprule
{$ \begin{array}{c|ccc}
 & \text{Rain Coat} & \text{Tee-shirt}  \\
\cline{1-3} 
\text{Rain Coat} & 0.5 &  0\\
\text{Tee-shirt} & 0 &  0.5\\
\end{array}$} & {$\begin{array}{c|cc}
 & \text{Rain Coat}  \\
\cline{1-2} 
\text{Rain Coat} & 0.5 \\
\text{Tee-shirt} & 0.5 \\
\end{array}$}\\
\bottomrule
\end{tabular}
\end{table}\\
In the table above, each row represents choice of an alternative in the first period and each column represents choice of an alternative in the second period. Notably, this is the same behavior as is induced by Example \ref{CorrelationConsDep}.
\end{example}

Examples \ref{CorrelationConsDep} and \ref{CorrelationStateDep} show that perfect correlation between first and second period choice can be achieved even while marginality holds. These two examples actually show something more. In Example \ref{CorrelationConsDep} we consider a transition function which is state independent and in Example \ref{CorrelationStateDep} we consider a transition function which is consumption independent (does not depend on the consumption input). This means that state dependence and consumption dependence can both independently induce perfectly correlated choice. As it turns out, we only need complete monotonicity and marginality to characterize consumption dependent random utility.

\begin{theorem}\label{CDRUM}
    A random joint choice rule $p$ is consistent with CDRUM if and only if it satisfies complete monotonicity and marginality.
\end{theorem}

Theorem \ref{CDRUM} offers a test for consumption dependent random utility when we have a rjcr. As mentioned prior, if we are instead faced with conditional choice probabilities, we can test complete monotonicity by simply transforming these conditional choice probabilities into a random joint choice rule. However, as marginality is the necessary and sufficient condition for the existence of a conditional choice probability representation\footnote{We have a conditional choice representation if and only if we can define first period choice independently of the second period menu. This gives us $p(x,A)$ in the first period and $p(y,B|x,A)$ in the second period.}, if we start with conditional choice probabilities, we no longer need to test marginality as it will automatically hold.

\subsubsection{Joint Choice vs Conditional Choice}

We now highlight the importance of our usage of random joint choice rules as our primitive. We begin by comparing our approach to that of \citet{frick2019dynamic}. We then show that, when conditional choice probabilities are our primitive and menus vary exogenously, it is difficult to distinguish between myopic and dynamically sophisticated agents.

\citet{frick2019dynamic} essentially consider a system of conditional choice probabilities.\footnote{Formally, \citet{frick2019dynamic} considers a data generating process where agents today choose their consumption today and their menu tomorrow. However, the characterizations from the paper remain largely unchanged when removing the menu choice component of the data generating process. See Chapter 7 of \citet{strzalecki2024stochastic} for a discussion of the model without menu choice.} \citet{frick2019dynamic} characterizes state dependent and consumption independent dynamic random expected utility as their main representation. In Appendix K of \citet{frick2017dynamic}, an earlier working paper version of \citet{frick2019dynamic}, the authors characterize a state and consumption dependent dynamic random expected utility model. Here the objects of choice are lotteries. Marginality, or an analogue thereof, does not appear in the axiomatization of either of these models. This is, at least in part, due to the usage of conditional choice probabilities in \citet{frick2019dynamic}, which already impose marginality on the underlying rjcr. However, as we have pointed out earlier, marginality, and thus having a well-defined system of conditional choice probabilities, is a behavioral consequence of the myopic nature of our agents.

To better highlight this point, we consider a version of our model with dynamic sophistication. Let $\mathcal{L}(X^2)$, with typical element $\rhd$, denote the set of preferences over alternative pairs. Let $\mu$ correspond to a typical distribution over elements of $\mathcal{L}(X^2)$.

\begin{definition}
    A random joint choice rule $p$ is \textbf{consistent} with the dynamic random utility model (DRUM) if there exists a probability distribution $\mu$ over preferences $\rhd$ such that the following holds for all $A \times B \subseteq X^2$ and $(x,y) \in A \times B$.
    \begin{equation}
        p(x,y,A,B) = \mu(\{\rhd|(x,y) \rhd (a,b) \text{ } \forall (a,b)\in A \times B\})
    \end{equation}
\end{definition}

DRUM is the extension of CDRUM that allows each agent to choose in the first period while taking into account the second period. Our next example shows that there are DRUM representations that fail to satisfy complete monotonicity and marginality.

\begin{example}[Intertemporal Complements]\label{CompExample}
Consider a single agent who is choosing over two period choice streams. The agent has a preference to consume both $x$ and $y$ in their consumption stream. Their preference is given by $(x,y)\rhd (y,x) \rhd (x,x) \rhd (y,y)$. The full choice probabilities given by this preference are in the following table.
        \begin{table}[h!]
    \centering
    \begin{tabular}{ccc}
\toprule
{$ \begin{array}{c|ccc}
 & x & y  \\
\cline{1-3} 
x & 0 &  1\\
y & 0 &  0\\
\end{array}$} & {$\begin{array}{c|cc}
 & x  \\
\cline{1-2} 
x & 0 \\
y & 1 \\
\end{array}$} & {$\begin{array}{c|cc}
 & y  \\
\cline{1-2} 
x & 1 \\
y & 0 \\
\end{array}$}\\
& & \\
{$ \begin{array}{c|ccc}
 & x & y  \\
\cline{1-3} 
x & 0 &  1\\
\end{array}$} & {$\begin{array}{c|cc}
 & x  \\
\cline{1-2} 
x & 1 \\
\end{array}$} & {$\begin{array}{c|cc}
 & y  \\
\cline{1-2} 
x & 1 \\
\end{array}$}\\
& & \\
{$ \begin{array}{c|ccc}
 & x & y  \\
\cline{1-3} 
y & 1 &  0\\
\end{array}$} & {$\begin{array}{c|cc}
 & x  \\
\cline{1-2} 
y & 1 \\
\end{array}$} & {$\begin{array}{c|cc}
 & y  \\
\cline{1-2} 
y & 1 \\
\end{array}$}\\
\bottomrule
\end{tabular}
\end{table}

We now observe that maximization of $\rhd$ induces failures of marginality. At choice set $\{x,y\} \times \{x,y\}$, our agent chooses $(x,y)$. However, when the choice set is given by $\{x,y\}\times \{x\}$, our agent chooses $(y,x)$, which means that first period choices are not independent of the second period choice set. Further, this random joint choice rule also fails to satisfy complete monotonicity. By directly applying the M\"{o}bius inverse formula, we find that $q(y,x,\{y\},\{x\})=-1$.

\end{example}

Example \ref{CompExample} shows that, by adding dynamic sophistication, DRUM can fail to satisfy the two characteristic axioms of CDRUM. However, in DRUM, every failure of complete monotonicity is tied to a failure of marginality.

\begin{theorem}\label{thm:dynamicsophistication}\footnote{Note that we only prove this theorem in the case of two periods. The proof extends to the general model, but the notation becomes overly cumbersome.}
    Suppose that a rjcr $p$ satisfies marginality and is consistent with DRUM. Then $p$ is consistent with CDRUM.
\end{theorem}

Theorem \ref{thm:dynamicsophistication} tells us that any behavioral difference between DRUM and CDRUM directly arises from failures of marginality in DRUM. This means that if a DRUM representation induces well-defined conditional choice probabilities, then it is behaviorally equivalent to some CDRUM representation with those same conditional choice probabilities. As such, conditional choice probabilities are not as well suited as random joint choice rules for distinguishing between myopic and dynamically sophisticated behavior.

\subsection{State Independence}
In the prior section we allowed for both consumption and state dependence. An analyst may instead want to consider an environment where there is no evolving state, thus ridding ourselves of state dependence, while continuing to allow for consumption dependence. As an example, an analyst may be interested to see if choices are driven only by habit formation, in which case the analyst may market and price their good in such a way to attract first time consumers. By restricting to rational choice but allowing for full heterogeneity in a population's preferences and consumption dependence, we arrive at the state independent consumption dependent random utility model. Our goal in this section is to characterize SI-CDRUM.

Before axiomatizing the SI-CDRUM, it is important to know if state independence implies any further restrictions over CDRUM. Example \ref{SIFurtherContent} shows that state independence has further empirical content beyond that CDRUM.

\begin{example}\label{SIFurtherContent}
    Consider the following choice probabilities.
\begin{table}[h!]
    \centering
    \begin{tabular}{cc}
\toprule
{$ \begin{array}{c|ccc}
 & x & y  \\
\cline{1-3} 
x & 0.5 &  0\\
y & 0 &  0.5\\
\end{array}$} & {$\begin{array}{c|ccc}
 & x & y  \\
\cline{1-3} 
y & 0.5 & 0\\
z & 0 & 0.5\\
\end{array}$}\\
\bottomrule
\end{tabular}
\end{table}\\
Above, each row row represents choice of an alternative in the first period and each column represents choice of an alternative in the second period. Consider an agent who chooses $y$ over $x$ in the first period. We see in the table that they choose $y$ over $x$ with certainty in the second period. Now consider the agent who chooses $y$ over $z$ in the first period. They choose $x$ over $y$ with certainty in the second period. However, in the case of state independence, the ranking over $x$ and $y$ in the second period should be the same when an agent chooses $y$ in the first period. Thus this behavior is inconsistent with state independence. Now suppose we are allowing for state dependence. If the first period choices are dictated by $x \succ y \succ z$, denoted $\succ_x$, half of the time and $z \succ y \succ x$, denoted $\succ_z$, half of the time, then the following state dependent transition function rationalizes the data.
\begin{equation*}
        \begin{split}
            t(\succ_x) = \begin{cases}
                1 \text{ } \succ_x \\
                0 \text{ } \succ_z
            \end{cases} & t(\succ_z) =\begin{cases}
                0 \text{ } \succ_x \\
                1 \text{ } \succ_z
            \end{cases} \\
        \end{split}
    \end{equation*}
\end{example}

Example \ref{SIFurtherContent} tells us that, conditional on choosing $y$ in the first period, the conditional probability of choosing $y$ over $x$ in the second period should be independent of our choice set in the first period. We can extend this logic beyond binary comparisons to say that that, conditional on choosing $x$ in the first period, choice probabilities in second period should be independent of the first period's choice set. Our next axiom captures exactly this.

\begin{axiom}[Choice Set Independence]\label{ChoiceSetInd}
    A random joint choice rule $p$ satisfies \textbf{choice set independence} if, for each $y \in B \subseteq X$, for each $(x,A)$ and $(x,A')$ with $p(x,A)>0$ and $p(x,A')>0$, we have $\frac{p(x,y,A,B)}{p(x,A)}=\frac{p(x,y,A',B)}{p(x,A')}$.
\end{axiom}

Notably, choice set independence puts no restriction on dependence of second period choice on first period choice. It simply restricts dependence on first period choice set. As it turns out, choice set independence is the only further restriction that state independence imposes over CDRUM.

\begin{theorem}\label{SICDRUMthm}
    A random joint choice rule $p$ is consistent with SI-CDRUM if and only if it satisfies complete monotonicity, marginality, and choice set independence.
\end{theorem}

We have already noted that state independence has further empirical content than the base model. Now we turn our attention to the case of consumption independence. This version of the model is studied in \citet{li2021axiomatization}, \citet{chambers2024correlated}, and \citet{kashaev2022nonparametric}. Consumption independence imposes a stronger form of marginality as a restriction on the data. Second period choices should be independent of the first period's choice set. Further, as Example \ref{SIFurtherContent} shows, there are consumption independent transition functions which can induce choice not consistent with consumption dependence. All of this together tells us that there are forms of consumption dependence which can not be mimicked by state dependence and there are forms of state dependence which can not be mimicked by consumption dependence.

\section{Revealed Preferences}\label{sec:hypo}

In the previous section, we provided an axiomatic characterization of the consumption dependent random utility model. In this section, we extend our discussion of this model and take a revealed preference style approach in order to build toward a test of CDRUM. In doing so, we drop the assumption that we observe choice on every possible product subset of $X^2$. Specifically, we assume that we observe choice probabilities for every product set contained in $\mathcal{X}_{lim}\subseteq \mathcal{X}^2$. We develop two existential linear programs which characterize CDRUM on a limited domain. The first is the analogue of the linear program used in \citet{kitamura2018nonparametric} to test the static random utility model. The second uses our Theorem \ref{CDRUM} and is the analogue of the linear program used in \citet{turansick2023alternative} to test static random utility. We then compare the computational complexity of the tests implied by these two linear programs as well as the min max complexity of the column generation procedure proposed in \citet{smeulders2021nonparametric} applied to our environment.

Up until now, we have thought of our initial distribution over preferences as separate from our transition function. We can instead think of treating these two objects as a single object. Consider a vector indexed by the elements of the form $(x,y,A,B)$ with $(x,y) \in A \times B$. We can use these vectors to encode deterministic choice. The deterministic analogue, in other words an extreme point, of CDRUM consists of a single linear order $\succ$ which encodes choice in the first period and a linear order $\succ_x$ for each $x \in X$ which encodes choice in the second period when our agent chooses $x$ in the first period. As such, if we fix a linear ordering of $X$, each element of $(\mathcal{L}(X))^{|X|+1}$ corresponds to one of these deterministic representations. Now consider a matrix $E$ whose rows are indexed by the elements of $(\mathcal{L}(X))^{|X|+1}$ and whose columns are indexed by elements of the form $(x,y,A,B)$ with $(x,y) \in A \times B$. In this matrix, the entry $e((\succ,\succ_{x_1},\dots,\succ_{x_{|X|}}),(x,y,A,B))$ is equal to one if $x = M(\succ,A)$ and $y = M(\succ_x,B)$ and zero otherwise. Each row of this $E$ matrix captures the choice pattern of one extreme point of the CDRUM. Notably, we could consider such an $E$ matrix even when we do not observe every choice set. We would then just remove each column indexed by $(\cdot,\cdot,A,B)$ if choice from $A \times B$ is unobserved. It is the case that, no matter which choice sets we observe, our data is consistent with CDRUM if and only if there exists some vector $r$ such that the following holds.
\begin{align}
    rE= p \label{ConicConsistency} \\
    r \geq 0 \label{ConicPos}
\end{align}

This linear program is the direct analogue in our setting to the linear program at the base of the test developed by \citet{kitamura2018nonparametric}. Unfortunately, as shown by \citet{smeulders2021nonparametric}, this test is computationally burdensome. Much of this computational burden comes from constructing the $E$ matrix used in Equation \ref{ConicConsistency}. As such, the test that Equations \ref{ConicConsistency} and \ref{ConicPos} build towards faces the same computational burdens in our environment. Our second test builds on the work of \citet{turansick2023alternative} and actually uses the characterization in Theorem \ref{CDRUM} to construct a less burdensome test. As shown in Theorem \ref{CDRUM}, CDRUM is characterized by complete monotonicity and marginality. Although the M\"{o}bius inverse of choice probabilities can itself be written as a function of the choice probabilities, it is important for our test that we are able to write down our marginality condition in terms of the M\"{o}bius inverse itself. To do this, we rely on a result from \citet{chambers2024correlated}.

\begin{lemma}[\citet{chambers2024correlated}]\label{recursivity2p}
    A random joint choice rule $p$ satisfies marginality if and only if for every $A \in \mathcal{X}$, every $B \subsetneq X$, and every $x\in A$ we have the following.
    \begin{equation}\label{recursivityEq}
        \sum_{y\in B} q(x,y,A,B) =\sum_{z \not \in B} q(x,z,A,B \cup\{z\})
    \end{equation}
\end{lemma}

Lemma \ref{recursivity2p} tells us that we can encode marginality directly in terms of the M\"{o}bius inverse of choice probabilities. It then follows that, if we observe choice for every possible choice set, CDRUM is characterized by a solution existing to the following linear program.
\begin{align}
    \sum_{A \subseteq A'} \sum_{B \subseteq B'}q(x,y,A',B') = p(x,y,A,B) \text{ }\forall A,B \in\mathcal{X}_{lim}, \forall (x,y) \in A\times B \label{Qconsistincy} \\
    \sum_{y\in B} q(x,y,A,B) =\sum_{z \not \in B} q(x,z,A,B \cup\{z\}) \text{ }\forall A \in \mathcal{X}, \forall B \subsetneq X, \forall x \in A \label{Qrecursivity} \\
    q(x,y,A,B) \geq 0 \text{ } \forall A,B \in \mathcal{X}, \forall (x,y) \in A \times B \label{Qpositivity}
\end{align}

Above, Equation \ref{Qconsistincy} encodes that our $q$ function is in fact the M\"{o}bius inverse of the observed choice probabilities. Equation \ref{Qrecursivity} encodes that the random joint choice rule satisfies marginality. Lastly, Equation \ref{Qpositivity} encodes that the random joint choice rule satisfies complete monotonicity. However, our goal is not to develop a test when we observe choice at every choice set, but rather to develop a test regardless of the choice sets we observe. The intuition behind how we handle limited data is that we essentially ask if there is an extension of our limited random joint choice rule to a full domain that satisfies Equations \ref{Qconsistincy}-\ref{Qpositivity}. Building on \citet{turansick2023alternative}, the trick here is that we can encode this extension problem in terms of the M\"{o}bius inverse of choice probabilities. Observe the following linear constraints.
\begin{align}
    \sum_{x \in A} \sum_{y \in X} q(x,y,A,X) = \sum_{z \not \in A} \sum_{y \in X} q(z,y,A\cup\{x\},X) \text{ } \forall A \subsetneq X \label{QinflowOutflow} \\
    \sum_{x \in X} \sum_{y \in X} q(x,y,X,X) = 1 \label{QInitialCon}
\end{align}

Recall the definition of random joint choice rules. They are characterized by three properties. The first property is that $\sum_{x\in A}\sum_{y\in B}p(x,y,A,B)$ does not depend on $A$ or $B$. When marginality holds, a weaker version of this property asks that $\sum_{x\in A}p(x,A)$ does not depend on $A$. Theorem 3.1 from \citet{turansick2023graphical} shows that this condition is equivalent to Equation \ref{QinflowOutflow}. The second property of random joint choice rules is that $\sum_{x\in A}\sum_{y\in B}p(x,y,A,B)$ is not only invariant but also sums to one. Again, a weaker version of this, when marginality holds, simply asks that $\sum_{x\in X}p(x,X)=1$, which is equivalent to Equation \ref{QInitialCon}. The last condition characterizing random joint choice rules is that they are non-negative. Equation \ref{Qpositivity} already implies non-negativity of our random joint choice rule. Thus far, we have said that Equations \ref{QinflowOutflow} and \ref{QInitialCon} are equivalent to conditions which are weaker than necessary to characterize random joint choice rules. It turns out that, when Equations \ref{Qrecursivity} and \ref{Qpositivity} hold, these weaker conditions are actually equivalent to the stronger conditions. Taking this all together, it follows that there exists a solution to the linear program described by Equations \ref{Qconsistincy}-\ref{QInitialCon} if and only if there exists a full domain rjcr which satisfies marginality, satisfies complete monotonicity, and is consistent with our observed rjcr. Our next theorem summarizes our discussion thus far.

\begin{theorem}\label{HypoTestLinProThm}
    The following are equivalent.
    \begin{enumerate}
        \item The (limited domain) random joint choice rule $p$ is consistent with CDRUM.
        \item There exists a vector $r$ which solves the linear program given by Equations \ref{ConicConsistency}-\ref{ConicPos}.
        \item There exists a vector $q$ which solves the linear program given by Equations \ref{Qconsistincy}-\ref{QInitialCon}.
    \end{enumerate}
\end{theorem}

At the beginning of this section, we mentioned that the test based on \citet{turansick2023alternative} offers computational improvements over the test based on \citet{kitamura2018nonparametric}. To show this, we need to make one further refinement of Equations \ref{Qconsistincy}-\ref{QInitialCon}. Recall that we use Equations \ref{QinflowOutflow}-\ref{QInitialCon} to ensure that our M\"{o}bius inverse $q$ induces choice probabilities for choice sets that we do not observe. As it turns out we only need to apply Equations \ref{QinflowOutflow} and \ref{QInitialCon} at the choice sets we do not observe. This is given by the following.

\begin{align}
    \sum_{x \in A} \sum_{y \in X} q(x,y,A,X) = \sum_{z \not \in A} \sum_{y \in X} q(z,y,A\cup\{x\},X) \text{ } \text{ if } \forall B \in \mathcal{X}, A \times B \not \in \mathcal{X}_{lim} \label{QinflowOutflowLim} \\
    \sum_{x \in X} \sum_{y \in X} q(x,y,X,X) = 1 \text{ if } \forall B \in \mathcal{X}, X \times B \not \in \mathcal{X}_{lim} \label{QInitialConLim}
\end{align}

The following theorem formalizes this equivalence.

\begin{theorem}\label{FasterHypoTest}
The following are equivalent.
    \begin{enumerate}
        \item The (limited domain) random joint choice rule $p$ is consistent with CDRUM.
        \item There exists a solution to Equations \ref{Qconsistincy}-\ref{QInitialCon}.
        \item There exists a solution to Equations \ref{Qconsistincy}-\ref{Qpositivity} and \ref{QinflowOutflowLim}-\ref{QInitialConLim}.
    \end{enumerate}
\end{theorem}

Thus far in this section, we have ignored the problem of finite data and have worked with idealized choice probabilities. Both of our tests, Equations \ref{ConicConsistency}-\ref{ConicPos} as well as Equations \ref{Qconsistincy}-\ref{Qpositivity} and \ref{QinflowOutflowLim}-\ref{QInitialConLim}, are tests which ask if there is a non-negative answer to a series of equality constraints. This is exactly the form studied in \citet{fang2023inference}. As such, both of these tests can be directly implemented as hypothesis tests using the technology of \citet{fang2023inference}.

\subsection{Computational Burden}

We are now ready to compare the computational burden of our two tests. Before proceeding, we note that Equations \ref{Qconsistincy}-\ref{Qpositivity} and \ref{QinflowOutflowLim}-\ref{QInitialConLim} can be represented in matrix form, $Fq=l$. Almost all of the computational burden of both of our tests come from constructing their respective matrices, $E$ in the case of Equations \ref{ConicConsistency}-\ref{ConicPos} and $F$ in the case of Equations \ref{Qconsistincy}-\ref{Qpositivity} and \ref{QinflowOutflowLim}-\ref{QInitialConLim}. As such, our focus will be on comparing the size of $E$ and $F$. As $F$ is at its largest when every choice set is observed, we make this comparison in the case where we observe every choice set. Observe that $E$ has the same number of columns as $F$. They each have one column for each $(x,y,A,B)$ with $(x,y) \in A\times B\in \mathcal{X}^2$. This means that we can simply compare the number of rows of each of these matrices. There is one row of $E$ for each extreme point of the consumption dependent random utility model. For two periods, the number of extreme points of the model is given by $|X|(|X|!)^2$.\footnote{To see this, there are $|X|!$ preferences which define first period choice. For each $x \in X$ and each $\succ \in \mathcal{L}(X)$, corresponding to first period choice and preference, we have $\succ'$ inducing second period choice. This corresponds to $|X|(|X|!)^2$ representations.} There is one row of $F$ for each constraint in Equations \ref{Qconsistincy} and \ref{Qrecursivity}. There is one constraint in Equation \ref{Qconsistincy} for each $(x,y,A,B)$ with $(x,y) \in A\times B$. This number is given by $(|X|2^{|X|-1})^2$. There is one constraint in Equation \ref{Qrecursivity} for each $(x,A,B)$ with $x \in A$ and $B \not \in \{\emptyset,X\}$. This number is given by $(|X|2^{|X|-1})(2^{|X|}-2)$. Taking these two together tells us that the total number of rows in $F$ is given by $(|X|2^{|X|-1})^2+|X|2^{2|X|-1}-|X|2^{|X|}$. It then follows that the number of rows in $E$ is larger than then number of rows in $F$ as long as $|X|\geq 4$. Table \ref{table:matrixsize} compares the number of rows of $E$ and $F$ for a few values of $|X|$.

\begin{table}[h!]
    \centering
    \begin{tabular}{|c|c|c|}
        \cline{1-3}
        $|X| $& $E$ rows & $F$ rows \\
        \cline{1-3}
        2 & $8$ & $24$ \\
        \cline{1-3}
        3 & $108$ & $216$\\
        \cline{1-3}
        4 & $2304$ & $1472$\\
        \cline{1-3}
        5 & $72,000$ & $8800$ \\
        \cline{1-3}
        6 &  $3,110,400$ & $48,768$ \\
        \cline{1-3}
        7 &  $177,811,200$ & $257,152$ \\
        \cline{1-3}
    \end{tabular}

    \caption{The number of rows in the $E$ and $F$ matrices are given as a function of $|X|$.}
    \label{table:matrixsize}
\end{table}

We now take a moment to discuss the column generation procedure of \citet{smeulders2021nonparametric}.\footnote{See also \citet{demuynck2022testing}.} When applied to our Equations \ref{ConicConsistency}-\ref{ConicPos}, the column generation approach begins by guessing that a certain collection of extreme points of CDRUM are not necessary to rationalize the data. In doing so, the value of $r_i$ is set to zero for each extreme point in this collection. The procedure then constructs a matrix $\Bar{E}$ that is the same as $E$ except it is missing the rows corresponding to each extreme point in our collection. This $\Bar{E}$ matrix forms an inner approximation of the cone formed by $E$. It then follows that, if our data is in the cone formed by $\Bar{E}$, then it is also in the cone formed by $E$. If our data does not lie in the cone formed by $\Bar{E}$, then we add some rows to $\Bar{E}$ corresponding to the extreme points in our original collection of excluded extreme points. \citet{smeulders2021nonparametric} develop a procedure to aid in the choice which extreme points to add to the matrix $\Bar{E}$.

We now compare the computational burden of applying this column generation procedure to our Equations \ref{ConicConsistency}-\ref{ConicPos} against the test given by Equations \ref{Qconsistincy}-\ref{Qpositivity} and \ref{QinflowOutflowLim}-\ref{QInitialConLim}. Since the construction of the $\Bar{E}$ matrix endogenously depends on the data, we compare the size of $F$ with the min max size of $\Bar{E}$. That is to say, suppose we first let nature choose some distribution over our extreme points and then we let our analyst construct $\Bar{E}$ given this distribution over extreme points. Nature's goal is to maximize the size of $\Bar{E}$ while the analyst's goal is to minimize $\Bar{E}$. In this case, the min max size of $\Bar{E}$ corresponds to at least the dimension of the vector space formed by our extreme points and at most the dimension of this vector space plus one (by Caratheodory's Theorem).

\begin{theorem}\label{dimThm}
    The dimension of the vector space spanned by the extreme points of CDRUM is given by $(|X|2^{|X|-1})^2-|X|2^{2|X|-1}+(|X|-1)2^{|X|}+2$.
\end{theorem}

Recall that the $F$ matrix has $(|X|2^{|X|-1})^2+|X|2^{2|X|-1}-|X|2^{|X|}$ rows. This means that the test given by Equations \ref{Qconsistincy}-\ref{Qpositivity} and \ref{QinflowOutflowLim}-\ref{QInitialConLim} is strictly more burdensome than the min max computational burden of the column generation procedure applied to Equations \ref{ConicConsistency}-\ref{ConicPos}. However, these two tests are both $\mathcal{O}((|X|2^{|X|-1})^2)$ and the test given by Equations \ref{Qconsistincy}-\ref{Qpositivity} and \ref{QinflowOutflowLim}-\ref{QInitialConLim} does not require ex ante knowledge of the distribution generating the data to reach $\mathcal{O}((|X|2^{|X|-1})^2)$ complexity. We end this section by noting that a similar column generation procedure can be applied to Equations \ref{Qconsistincy}-\ref{Qpositivity} and \ref{QinflowOutflowLim}-\ref{QInitialConLim} when not every choice set is observed.

\section{Conclusion}\label{sec:conc}

In this paper we consider a model of dynamic random utility that allows for an agent's preference to depend on their history of choice. We axiomatically analyze this model and provide the first characterization of dynamic random utility in terms of a finite set of linear inequalities. Using this axiomatization, we are able to develop a finite existential linear program that characterizes consumption dependent random utility on an arbitrary domain. This linear program can be tested with real data using the techniques of \citet{fang2023inference}.

A key distinction important to our analysis is that between \textit{history dependence}, the statistical dependence of today's choice on yesterday's choice, state dependence, the dependence of today's utility on a dynamically evolving \textit{exogenous} state, and consumption dependence, the dependence of today's preference on yesterday's choice. Although we follow the language of \citet{frick2019dynamic}, the distinction between history dependence and consumption dependence shows up in \citet{heckman1978simple} and \citet{heckman1981heterogeneity}. Under the assumption of preference maximization, we find that there are types of history dependence that cannot arise even in the presence of consumption and state dependence. Behaviors such as the mere-exposure effect are ruled out by our complete monotonicity axiom, which is an extension of the classic static random utility axiom to a dynamic environment.\footnote{Notably, the mere-exposure effect is also ruled out by the dynamic random expected utility model of \citet{frick2019dynamic}. Our finding implies that the mere-exposure effect is not only ruled out by dynamic random \textit{expected} utility, but more generally by dynamic random utility.} Additionally, marginality rules out the statistical dependence of today's choice on an agent's future menus. The marginality axiom is independent of our assumption of preference maximization. This means that data generated by a consumption and state dependent but ``irrational" process will still satisfy marginality. In the context of dynamic choice, this means that, if failures of marginality are observed, they arise due to \textit{dynamic sophistication}, the ability of an agent to take into account the impact of their choice today on their utility in future periods.

\subsection{Related Literature}

We now conclude with a discussion of the related literature. Our paper builds on the literature which axiomatically studies dynamic random utility. This literature begins with \citet{fudenberg2015dynamic} which studies the dynamic logit model typically used in dynamic discrete choice settings. Our paper is more closely related to \citet{frick2019dynamic} which, to our knowledge, is the first to axiomatize a general nonparametric model of dynamic random utility. They study an extension of the random expected utility model of \citet{gul2006random} to a dynamic environment. Their base representation allows for state dependence but not consumption dependence. The goal of their axiomatization is to characterize exactly which forms of history dependence can arise even in the absence of consumption dependence. In the appendix of \citet{frick2017dynamic}, a working version of \citet{frick2019dynamic}, they consider an extension of their model which allows for consumption dependence. However, their axiomatization relies on the linear nature of expected utility and the rich domain induced by choice over lotteries. We work with general linear orders over a finite set of abstract alternatives, so the axioms of \citet{frick2019dynamic} lose much of their meaning in our setting. \citet{duraj2018dynamic} is able to extend many of the results of \citet{frick2019dynamic} to a setting with an objective state space.

More recently, there has been work by \citet{li2021axiomatization}, \citet{chambers2024correlated}, and \citet{kashaev2022nonparametric} which studies agents who are subject to state dependence in a general abstract setting. The models of these papers correspond to the special case of our model when the underlying transition function is consumption independent. Each of these papers show that complete monotonicity and a stronger form of marginality are necessary for their representation. However, it is still an open question as to how to axiomatize these models in terms of a finite set of linear inequalities as we do in our Theorem \ref{CDRUM}. \citet{li2021axiomatization} is able to offer a full (infinite) characterization of the model in the style of \citet{clark1996random}. \citet{chambers2024correlated} focuses much of their work on studying the marginality axiom and its relationship to separable utility. A large part of \citet{kashaev2022nonparametric} focuses on extending the techniques of \citet{kitamura2018nonparametric} to a dynamic environment.

Beyond the work we have already mentioned, \citet{lu2018random} study intertemporal choice when the agent's discount rate is random. \citet{pennesi2021intertemporal} studies the difference between intertemporal Luce an logit models. The key difference between the two models is that the discount factor enters exponentially in the logit model while it does not in the Luce model. \citet{strack2021dynamic} consider a two period model. In the second period, the agent chooses according to a random utility. In the first period, the agent chooses according to an expected utility function. \citet{strack2021dynamic} study when the expected utility function in the first period can be induced by the random utility in the second period. \citet{deb2021dynamic} study a model of common learning. In this model, they study when the dynamic choices of a population of expected utility maximizers can be induced by a common stream of information.

More generally, we contribute to the literature studying random utility. The model was first introduced by \citet{block1959random} and first axiomatized by \citet{falmagne1978representation}. From a technical aspect, our proof technique builds on the work of \citet{fiorini2004short} and \citet{chambers2024correlated}. \citet{fiorini2004short} offers a proof of the characterization of \citet{falmagne1978representation} using graph theoretic techniques. \citet{chambers2024correlated} extends the graphical representation considered in \citet{fiorini2004short} to allow for choice over multiple agents/periods. We refine the graphical representation used in \citet{chambers2024correlated} and use it to prove our axiomatization. Recently, \citet{kono2023axiomatization} uses similar graphical techniques to those of \citet{fiorini2004short} to study the empirical content of an outside option in the random utility model when the outside option is actually multiple goods. \citet{apesteguia2017single} studies a variant of the random utility model under an ordered domain assumption. As an alternative approach to \citet{falmagne1978representation}, \citet{mcfadden1990stochastic} offers a characterization of random utility through the theorem of the alternative. \citet{kitamura2018nonparametric} build on the intuition of this result in order to construct a hypothesis test of the random utility model. \citet{turansick2023alternative} offers a characterization of random choice rules through their M\"{o}bius inverse. As an application of this, \citet{turansick2023alternative} develops an alternative hypothesis test which is computationally less burdensome than the one of \citet{kitamura2018nonparametric}. We build on the work of \citet{kitamura2018nonparametric} and \citet{turansick2023alternative} by extending their techniques to the consumption dependent random utility model.

\appendix

\section{Extended Model and Graphical Construction}\label{extendedmodel}
In this appendix, we first consider an extension of the model we consider in Section \ref{Model} and then provide a graphical construction that is used to prove the results stated in Section \ref{sec:axioms}. Specifically, we extend the model beyond two periods to an arbitrary but finite number of periods. There are $T$ periods and we let $\tau \in \{1,\dots,T\}$ denote a specific time period. We use $\mathbf{x}$ to denote a vector of alternatives and $\mathbf{A}$ to denote a product of choice sets. Further, we use $\mathbf{x}^\tau$ to denote a vector of alternatives of length $\tau$. Specifically, $\mathbf{x}^\tau$ denotes the choices from periods $1$ through $\tau$. Similarly, we use $\mathbf{A}^\tau$ to denote the product of the choice sets of the first $\tau$ periods. Finally, we use $x_\tau$ and $A_\tau$ to denote the choice and choice set, respectively, in period $\tau$. We study random joint choice rules of the form $p(\mathbf{x}^T,\mathbf{A}^T)$. For a random joint choice rule $p$, we can define the M\"{o}bius inverse, $q$, of $p$ as follows.

\begin{equation}\label{mobiusfull}
    p(\mathbf{x}^T,\mathbf{A}^T)=\sum_{A_1 \subseteq A'_1}\cdots \sum_{A_T \subseteq A'_T}q(\mathbf{x}^T,\mathbf{A'}^T)
\end{equation}

For a random joint choice rule $p$ satisfying marginality (see Axiom \ref{marginalitygen} for the full version), $p(\mathbf{x}^\tau,\mathbf{A}^\tau)$ is well-defined for each $\tau \in \{1,\dots,T\}$. As such, we can consider the M\"{o}bius inverse of each of these random joint choice rules.

\begin{equation}\label{mobiusmargfull}
    p(\mathbf{x}^\tau,\mathbf{A}^\tau)=\sum_{A_1 \subseteq A'_1}\cdots \sum_{A_\tau \subseteq A'_\tau}q(\mathbf{x}^\tau,A'_1,\dots,A'_\tau)
\end{equation}

In Section \ref{Model}, we considered a two period model and so there was no difference between agents with Markovian consumption dependence and longer form of consumption dependence. The extended model we consider allows for consumption dependence of arbitrary finite length. As such, we need to consider an extension of transition functions.

\begin{definition}
    We call a function $t^n:X^n \times \mathcal{L}(x)^n \rightarrow \Delta(\mathcal{L}(X))$ a \textbf{transition function} of degree $n$.
\end{definition}

A transition function of degree $n$ allows for dependence on consumption and state histories of up to length $n$. The proofs of our results on random joint choice rules rely on a specific graphical construction. This graphical construction is an extension of the one considered in \citet{fiorini2004short} and is a special case of the graphical construction considered in \citet{chambers2024correlated}. Simply, our graphical construction constructs one graph of the form considered in \citet{fiorini2004short} for each history of choices (including the null history). For a history of form $(\mathbf{x}^\tau,x_{\tau+1},\mathbf{A}^\tau,A_{\tau+1})$ with $\tau<T$, our graphical representation captures, for agents with this history of choice, the choices not already explained by agents with the history $(\mathbf{x}^\tau,x_{\tau+1},\mathbf{A}^\tau,B_{\tau+1})$ for all possible choices of $B_{\tau+1}\neq A_{\tau+1}$.

We now present the graph corresponding to history $(\mathbf{x}^\tau,\mathbf{A}^\tau)$. The graph for each other history is constructed analogously. To begin, our graph has a node for each element of $2^X$, the power set of $X$. We index these nodes by the elements of $2^X$ and will refer to nodes by their index. There exists an edge between nodes $A$ and $B$ if one of the following two conditions hold.
\begin{enumerate}
    \item $A \subseteq B$ and $|B \setminus A| = 1$
    \item $B \subseteq A$ and $|A \setminus B| = 1$
\end{enumerate}
We assign each edge an edge capacity. Recall from Equation \ref{mobiusmargfull}, for a random joint choice rule satisfying marginality, we can take the M\"{o}bius inverse of random joint choice rule truncated to any history. As such, we can consider the M\"{o}bius inverse of $p(\mathbf{x}^\tau,x_{\tau+1},\mathbf{A}^\tau,A_{\tau+1})$ for each choice of $x_{t\tau+1}\in A_{\tau+1}\subseteq X$. We denote this M\"{o}bius inverse as $q(\mathbf{x}^\tau,x_{\tau+1},\mathbf{A}^\tau,A_{\tau+1})$. For the edge connecting sets $B$ and $B \setminus \{y\}$, we assign $q(\mathbf{x}^\tau,y_{\tau+1},\mathbf{A}^\tau,B_{\tau+1})$ as the edge capacity.

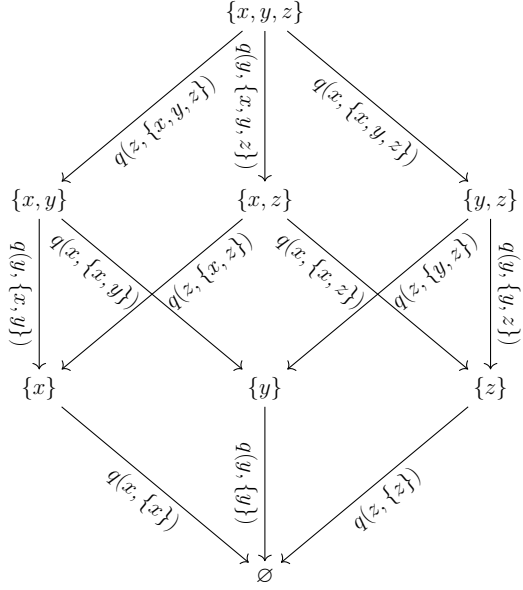
\begin{figure}
    \centering
\begin{tikzpicture}[scale=.5, transform shape]
    \Large
    \tikzstyle{every node} = [rectangle]
    
        \node (a) at (6,0) {$\emptyset$};
        
        \node (b) at (0,5) {$\{x\}$};
        \node (c) at (6,5) {$\{y\}$};
        \node (d) at (12,5) {$\{z\}$};
        
        \node (e) at (0,10) {$\{x,y\} $};
        \node (f) at (6,10) {$\{x,z\}$};
        \node (g) at (12,10) {$\{y,z\}$};

        \node (h) at (6,15) {$\{x,y,z\}$};
        
        \draw [->] (h) -- (f) node[midway, below, sloped] {$q(y,\{x,y,z\})$};
        \draw [->] (h) -- (g) node[midway, below, sloped] {$q(x,\{x,y,z\})$};  
        \draw [->] (h) -- (e) node[midway, below, sloped] {$q(z,\{x,y,z\})$};

        \draw [->] (e) -- (b) node[midway, below, sloped] {$q(y,\{x,y\})$}; 
        \draw [->] (e) -- (c) node[pos=.25, below, sloped] {$q(x,\{x,y\})$};
        \draw [->] (f) -- (b) node[pos=.25, below, sloped] {$q(z,\{x,z\})$};
        \draw [->] (f) -- (d) node[pos=.25, below, sloped] {$q(x,\{x,z\})$};
        \draw [->] (g) -- (c) node[pos=.25, below, sloped] {$q(z,\{y,z\})$};
        \draw [->] (g) -- (d) node[midway, above, sloped] {$q(y,\{y,z\})$}; 

        \draw [->] (b) -- (a) node[midway, below, sloped] {$q(x,\{x\})$};
        \draw [->] (c) -- (a) node[midway, below, sloped] {$q(y,\{y\})$};
        \draw [->] (d) -- (a) node[midway, below, sloped] {$q(z,\{z\})$};
    
    \end{tikzpicture}
    \caption{The graphical representation of $p$ for the null history and for the set $X = \{a,b,c\}$.}
    \label{fig:flowdiagram}
\end{figure}

We are interested in representing linear orders using our graphs. In each of these graphs, we can think of a path from $X$ to $\emptyset$ as a finite sequence of sets $\{A_i\}_{i=0}^{|X|}$ with $A_i \supsetneq A_{i+1}$. Each path on this graph is bijectively associated with a linear order. Specifically, we have that $x\succ y \succ z \succ \dots$ is associated with the path $X \rightarrow X \setminus \{x\}\rightarrow X \setminus \{x,y\}\rightarrow X \setminus \{x,y,z\} \rightarrow \dots$. This bijective association is due to our Theorem \ref{Uniqueness}. For data that is consistent with consumption dependent random utility, each $q(x_1,A_1)$ corresponds to the probability weight $\nu$ puts on drawing a preference which chooses $x_1$ from $A_1$ but does not choose $x_1$ from any superset of $A_1$. Each path on our graph corresponds to a collection of sets of preferences. The intersection of each of these sets leaves of us with a single linear order, the one associated with the path. This logic and bijective association extends analogously beyond the null history. Figure \ref{fig:flowdiagram} offers an example of this graph for the null history when $X=\{x,y,z\}$.

\section{Proofs from Section \ref{sec:axioms}}\label{app:axioms}

The results in Section \ref{sec:axioms} can be extended to any finite number of periods using the extended model presented in Appendix \ref{extendedmodel}. 

\begin{definition}
    A random joint choice rule $p$ is \textbf{consistent} with consumption dependent random utility if there exists a probability distribution over preferences $\nu$ and a series of transition functions $\{t^1,\dots,t^{T-1}\}$ such that the following holds.
    \begin{equation}\label{cdrumfulldef}
        p(\mathbf{x}^T,\mathbf{A}^T)=\sum_{\succ_1 \in N(x_1,A_1)} \cdots \sum_{\succ_T \in N(x_T,A_T)} \nu(\succ_1) \prod_{\tau=2}^T t^{\tau-1}_{\succ_{\tau}}(\mathbf{x}^{\tau-1},\succ^{\tau-1}) 
    \end{equation}
\end{definition}
In Equation \ref{cdrumfulldef} we use $\succ^\tau$ to denote a vector of linear orders of length $\tau$.

\subsection{Proof of Theorem \ref{Uniqueness}}
Before proving Theorem \ref{Uniqueness}, we extend it to any finite number of periods using the model presented in Appendix \ref{extendedmodel}. We then prove the extended result. Recall that $I(x,A)$ denotes the set of preferences which choose $x$ from $A$ but fail to choose $x$ from any choice set $A \cup \{z\}$. 

\begin{theorem}\label{uniquenessfullthm}
    A distribution over preferences $\nu$ and a sequence of transition functions $\{t^1,\dots,t^{T-1}\}$ are a consumption dependent random utility representation for random joint choice rule $p$ if and only if the following holds for all $\mathbf{A}^T \in \mathcal{X}^T$ and for each $\mathbf{x}^T \in \mathbf{A}^T$.
    \begin{equation}\label{uniquenessfull}
        q(x^T,A^T)=\sum_{\succ_1 \in I(x_1,A_1)} \cdots \sum_{\succ_T \in I(x_T,A_T)} \nu(\succ) \prod_{\tau=2}^T t^{\tau-1}_{\succ_{\tau}}(\mathbf{x}^{\tau-1},\succ^{\tau-1})
    \end{equation}
\end{theorem}

\begin{proof}
    Suppose that a distribution over preferences $\nu$ and a sequence of transition functions $\{t^1,\dots,t^{T-1}\}$ are a consumption dependent random utility representation for random joint choice rule $p$. Then each $p(\mathbf{x}^T,\mathbf{A}^T)$ can be written as follows.
    \begin{equation*}
        \begin{split}
            p(\mathbf{x}^T,\mathbf{A}^T) &=\sum_{\succ_1 \in N(x_1,A_1)} \cdots \sum_{\succ_T \in N(x_T,A_T)} \nu(\succ) \prod_{\tau=2}^T t^{\tau-1}_{\succ_{\tau}}(\mathbf{x}^{\tau-1},\succ^{\tau-1}) \\
            & = \sum_{A_1 \subseteq A'_1}\cdots \sum_{A_T\subseteq A'_T}\left(\sum_{\succ_1 \in I(x_1,A'_1)} \cdots \sum_{\succ_T \in I(x_T,A'_T)} \nu(\succ) \prod_{\tau=2}^T t^{\tau-1}_{\succ_{\tau}}(\mathbf{x}^{\tau-1},\succ^{\tau-1})\right) \\
            & = \sum_{A_1 \subseteq A'_1}\cdots \sum_{A_T\subseteq A'_T}q(\mathbf{x}^T,\mathbf{A'}^T)
        \end{split}
    \end{equation*}
    The equality in the first line follows from the definition of consumption dependent random utility. The equality in the second line follows from the observation that $N(x,A)=\bigcup_{A \subseteq A'}I(x,A')$ with each of these $I(x,A')$ being disjoint. The equality in the last line follows from the definition of the M\"{o}bius inverse of $p(\mathbf{x}^T,\mathbf{A}^T)$. It then immediately follows that Equation \ref{uniquenessfull} holds for all $\mathbf{A}^T \in \mathcal{X}^T$ and for each $\mathbf{x}^T \in \mathbf{A}^T$ and so we are done.
\end{proof}

\subsection{Proof of Theorem \ref{CDRUM}}

We first extend the axioms used in Theorem \ref{CDRUM} to versions that accommodate the extended model discussed in Appendix \ref{extendedmodel}.
\begin{axiom}[Marginality]\label{marginalitygen}
    A random joint choice rule $p$ satisfies \textbf{marginality} if for every $\tau \in \{1,\dots,T-1\}$, every $\mathbf{A}^\tau \in \mathcal{X}^t$, every $\mathbf{x}^\tau \in \mathbf{A}^\tau$, and for every $B,C \in \mathcal{X}$ we have the following.
    \begin{equation}
        \sum_{y_{\tau+1} \in B_{\tau+1}}p(\mathbf{x}^\tau,y_{\tau+1},\mathbf{A}^\tau,B_{\tau+1})=\sum_{y_{\tau+1} \in C_{\tau+1}}p(\mathbf{x}^\tau,y_{\tau+1},\mathbf{A}^\tau,C_{\tau+1})
    \end{equation}
\end{axiom}

When marginality is satisfied, $p(\mathbf{x}^\tau,A^\tau)$ is well-defined and thus we can use this notation to discuss choice up to period $\tau$.

\begin{axiom}[Complete Monotonicity]\label{compmongen}
    A random joint choice rule $p$ is \textbf{completely monotone} if for every $\mathbf{A}^T \in \mathcal{X}^T$ and every $\mathbf{x}^T \in \mathbf{A}^T$, we have $q(\mathbf{x}^T,\mathbf{A}^T) \geq 0$.
\end{axiom}

We now restate Theorem \ref{CDRUM} in its full generality.

\begin{theorem}\label{cdrumfull}
    A random joint choice rule $p$ is consistent with consumption dependent random utility if and only if it satisfies Axioms \ref{marginalitygen} and \ref{compmongen}.
\end{theorem}

Note that Theorem \ref{CDRUM} is a special case of Theorem \ref{cdrumfull} in which $T=2$. Before proceeding to the proof of Theorem \ref{cdrumfull}, we first state a necessary preliminary result. This result first appears as Lemma 2 in \citet{chambers2024correlated} and we alter it slightly to fit our notation.

\begin{lemma}[\citet{chambers2024correlated}]\label{recursivity}
    A random joint choice rule $p$ satisfies marginality if and only if for each $\tau \in \{1,\dots,T-1\}$, every $\mathbf{A}^\tau \in \mathcal{X}^t$, every $\mathbf{x}^\tau \in \mathbf{A}^\tau$, and for every $B \subsetneq {X}$ we have the following.
    \begin{equation}\label{recursivityFullEq}
        \sum_{y_{\tau+1} \in B_{\tau+1}} q(\mathbf{x}^\tau,y_{\tau+1},\mathbf{A}^\tau,B_{\tau+1}) =\sum_{z_{\tau+1} \not \in B_{\tau+1}} q(\mathbf{x}^\tau,z_{\tau+1},\mathbf{A}^\tau,B_{\tau+1}\cup\{z_{\tau+1}\})
    \end{equation}
\end{lemma}

We make one further observation before we begin our proof. When a random joint choice rule $p$ satisfies marginality, $p(x_1,A_1)$ is well defined and thus choices from the first period can be thought of a classic random choice rule. The following result from \citet{falmagne1978representation} puts restrictions on the M\"{o}bius inverse of random choice rules.

\begin{theorem}[\citet{falmagne1978representation}]\label{Falmagneinflow}
    A random choice rule $p$ satisfies the following for all $\emptyset \subsetneq A_1 \subsetneq X$.
    \begin{equation}
        \sum_{x \in A}q(x_1,A_1) = \sum_{z_1 \not \in A_1}q(z_1,A_1 \cup \{z_1\})
    \end{equation}
\end{theorem}

Lemma \ref{recursivity} and Theorem \ref{Falmagneinflow} will be important in the proof of Theorem \ref{cdrumfull}. Notably, we are able to interpret these two results in terms of our graphical construction from Appendix \ref{extendedmodel}. These results tell us that at every interior node of every graph in our graphical construction, the total inflow due to edge capacities is equal to the total outflow due to edge capacities. We now proceed with our proof of Theorem \ref{cdrumfull}.

\begin{proof}
    We begin with necessity. By Theorem \ref{uniquenessfullthm}, we know that $q(\mathbf{x}^T,\mathbf{A}^T)$ corresponds to the probability weight put on a set of linear orders and thus must be non-negative. Thus Axiom \ref{compmongen} holds. To prove that Axiom \ref{marginalitygen} is necessary, recall the definition of a consumption dependent random utility representation.
    \begin{equation*}
        p(\mathbf{x}^T,\mathbf{A}^T)=\sum_{\succ_1 \in N(x_1,A_1)} \cdots \sum_{\succ_T \in N(x_T,A_T)} \nu(\succ) \prod_{\tau=2}^T t^{\tau-1}_{\succ_{\tau}}(\mathbf{x}^{\tau-1},\succ^{\tau-1}) 
    \end{equation*}
    Notably, if we have a consumption dependent random utility representation for $T$ periods, then we have a consumption dependent random utility representation for $T-1$ periods. This gives us the following.
        \begin{equation*}
        p(\mathbf{x}^{T-1},\mathbf{A}^{T-1})=\sum_{\succ_1 \in N(x_1,A_1)} \cdots \sum_{\succ_{T-1} \in N(x_{T-1},A_{T-1})} \nu(\succ) \prod_{\tau=2}^{T-1} t^{\tau-1}_{\succ_{\tau}}(\mathbf{x}^{\tau-1},\succ^{\tau-1}) 
    \end{equation*}
    Observe that $p(\mathbf{x}^{T-1},\mathbf{A}^{T-1})$ can be written independently of the choice set in period $T$. This shows that Axiom \ref{marginalitygen} holds.

    We now show sufficiency of Axioms \ref{marginalitygen} and \ref{compmongen}. Our proof proceeds as follows. We first use the graphical construction in Appendix \ref{extendedmodel} to represent the random joint choice rule. Then we show that this graphical representation can be completely decomposed into a series of path flows on each graph. We conclude by showing how these path flows can be translated into the primitives of our model. We begin by making three observations.
    \begin{enumerate}
        \item When Axiom \ref{marginalitygen} holds, each graph in our graphical construction satisfies inflow equals outflow at each interior node of the graph. This is shown through Lemma \ref{recursivity} and Theorem \ref{Falmagneinflow} and the discussion prior to this proof.
        \item When Axiom \ref{marginalitygen} holds, $q(\mathbf{x}^\tau,\mathbf{A}^\tau)= \sum_{y_{\tau+1}\in X_{\tau+1}} q(\mathbf{x}^\tau,y_{\tau+1},\mathbf{A}^\tau,X_{\tau+1})$. This follows from the following.
        
        \begin{equation*}
        \begin{split}
            p(\mathbf{x}^\tau,\mathbf{A}^\tau) & = \sum_{y_{\tau+1}\in X_{\tau+1}} p(\mathbf{x}^\tau,y_{\tau+1},\mathbf{A}^\tau,X_{\tau+1}) \\
            & =  \sum_{A_1 \subseteq A'_1}\cdots \sum_{A_\tau \subseteq A'_\tau} \sum_{y_{\tau+1}\in X_{\tau+1}} q(\mathbf{x}^\tau,y_{\tau+1},\mathbf{A'}^\tau,X_{\tau+1})\\
            p(\mathbf{x}^\tau,\mathbf{A}^\tau)& =  \sum_{A_1 \subseteq A'_1}\cdots \sum_{A_\tau \subseteq A'_\tau} q(\mathbf{x}^\tau\mathbf{A'}^\tau)
        \end{split}
        \end{equation*}
        Above, the first equality holds due to Axiom \ref{marginalitygen}. The second and third equalities hold due to the definition of the M\"{o}bius inverse. It then immediately follows that $q(\mathbf{x}^\tau,\mathbf{A}^\tau)= \sum_{y_{\tau+1}\in X_{\tau+1}} q(\mathbf{x}^\tau,y_{\tau+1},\mathbf{A}^\tau,X_{\tau+1})$.
        
        \item When Axioms \ref{marginalitygen} and \ref{compmongen} hold, each edge capacity is non-negative. This follows as each edge capacity in our graphical construction can be written as either $q(\mathbf{x}^T,\mathbf{A}^T)$ or as $q(\mathbf{x}^\tau,\mathbf{A}^\tau)= \sum_{y_{\tau+1}\in X_{\tau+1}} q(\mathbf{x}^\tau,y_{\tau+1},\mathbf{A}^\tau,X_{\tau+1})$.
    \end{enumerate}

    We now construct an algorithm which takes as input one of the graphs in our graphical construction. This algorithm takes this graph and constructs a path flow decomposition of this graph. Using the bijective association between paths and linear orders, this path flow decomposition is equivalent to a probability distribution over linear orders.

    \begin{algorithm}\label{decompalg}
    Take as input the graph associated with the potentially null history $(\mathbf{x}^\tau,\mathbf{A}^\tau)$.
        \begin{enumerate}
            \item If $q(\mathbf{x}^{\tau},x_{\tau+1},\mathbf{A}^\tau,X_{\tau+1})=0$ for all $x_{\tau+1} \in X_{\tau+1}$ set $\nu_{(\mathbf{x}^\tau,\mathbf{A}^\tau)}(\cdot)$ as the uniform distribution and terminate the algorithm. If not, then proceed to the step 2.
            \item Initialize at $i=0$, $\nu_{(\mathbf{x}^\tau,\mathbf{A}^\tau)}(\cdot)=0$, and $q_i(\mathbf{x}^{\tau},x_{\tau+1},\mathbf{A}^\tau,A_{\tau+1})=\\q(\mathbf{x}^{\tau},x_{\tau+1},\mathbf{A}^\tau,A_{\tau+1})\bigg/\left( \sum_{y_{\tau+1}\in X_{\tau+1}}q(\mathbf{x}^{\tau},y_{\tau+1},\mathbf{A}^\tau,X_{\tau+1}) \right)$. By step 1, the denominator is non-zero. By complete monotonicity, the denominator is positive. By construction, $\sum_{x_{\tau+1}\in X_{\tau+1}}q_i(\mathbf{x}^{\tau},x_{\tau+1},\mathbf{A}^\tau,X_{\tau+1})=1$.
            \item As there is some $q_i(\mathbf{x}^{\tau},x_{\tau+1},\mathbf{A}^\tau,A_{\tau+1})>0$, it follows from our first observation above that there is some path from $X$ to $\emptyset$ such that each edge has strictly positive edge capacity along this path. Fix this path. Set $r_i$ equal to the minimum edge capacity along this path. Let $\succ_i$ be the preference bijectively associated with this path. Set $\nu_{(\mathbf{x}^\tau,\mathbf{A}^\tau)}(\succ_i)=r_i$. For each $q_i(\mathbf{x}^{\tau},x_{\tau+1},\mathbf{A}^\tau,A_{\tau+1})$ associated with an edge of our chosen path, set $q_i(\mathbf{x}^{\tau},x_{\tau+1},\mathbf{A}^\tau,A_{\tau+1})=q_i(\mathbf{x}^{\tau},x_{\tau+1},\mathbf{A}^\tau,A_{\tau+1})-r_i$. For all other edges, set $q_i(\mathbf{x}^{\tau},x_{\tau+1},\mathbf{A}^\tau,A_{\tau+1})=q_i(\mathbf{x}^{\tau},x_{\tau+1},\mathbf{A}^\tau,A_{\tau+1})$. If there exists no $q_i(\mathbf{x}^{\tau},x_{\tau+1},\mathbf{A}^\tau,A_{\tau+1})>0$, terminate the algorithm. Otherwise, set $i=i+1$ and return to the start of step 3.
        \end{enumerate}
    \end{algorithm}

    We now make some claims about this algorithm. Note that at every step of the algorithm, we are subtracting out some constant weight along a fixed path. By doing so, we maintain the inflow equals outflow property from our first observation at every step of the algorithm. Second, as we are subtracting out the minimum edge capacity along our chosen path at every step of the algorithm and because each edge capacity is non-negative at initialization, at every step of the algorithm, our edge weights remain non-negative. Third, because of the inflow equals outflow property, the observation we make at the start of step 3 holds at every step of the algorithm. Notably, if we have some strictly positive $q_i(x,A)$, then there is some strictly positive $q(y,A\setminus\{x\})$ if $A \neq \emptyset$ and some strictly positive $q(z,A \cup\{z\})$ if $A \neq X$. This logic can be continued to show that there exists a full path from $X$ to $\emptyset$ with strictly positive edge capacities. Lastly, as there are a finite number of edges in our graph, our algorithm terminates in finite time and terminates with zero edge capacity everywhere on the graph. If there were some strictly positive edge capacity, then by the prior logic, we could find a strictly positive path from $X$ to $\emptyset$ and our algorithm would proceed with one more iteration. As the total outflow from $X$ at the start of the algorithm is equal to one (see step 2 of the algorithm), this means that the total weight assigned to $\nu_{(\mathbf{x}^\tau,\mathbf{A}^\tau)}(\cdot)$ is equal to one and is everywhere non-negative.

    We now have a series of distributions of linear orders. Our goal now is to assign these distributions to the distribution and transition functions in our representation in order to guarantee that our representation is consistent with the data. For $\nu_{\emptyset}(\cdot)$, the distribution for the null history, set $\nu(\cdot)=\nu_{\emptyset}(\cdot)$, where $\nu(\cdot)$ is the distribution over preferences in the first period. Our next step is to assign our constructed probability distributions to the outputs of transition functions. We will do this by assigning our distributions to the outputs of all preferences which fall in the set $I(x,A)$. As such, we use $t(x,I(x,A))$ to denote the transition function when $(x,\succ)$ is the input for any preference $\succ \in I(x,A)$. Given history $(\mathbf{x}^\tau,\mathbf{A}^\tau)$, assign $t_\succ^\tau(\mathbf{x}^\tau,I(x_1,A_1),\dots,I(x_\tau,A_\tau))=\nu_{(\mathbf{x}^\tau,\mathbf{A}^\tau)}(\succ)$. We now verify that our representation is consistent with the observed data. There are two cases. The first case is when $q(\mathbf{x}^T,\mathbf{A}^T)>0$.
    \begin{equation}\label{verification}
        \begin{split}
            & \sum_{\succ_1 \in I(x_1,A_1)} \cdots \sum_{\succ_T \in I(x_T,A_T)} \nu(\succ) \prod_{\tau=2}^T t^{\tau-1}_{\succ_{\tau}}(\mathbf{x}^{\tau-1},\succ^{\tau-1}) \\
           & = q(x_1,A_1) \prod_{\tau=2}^T t^{\tau-1}_{\succ_{\tau}}\frac{q(\mathbf{x}^{\tau-1},x_{\tau},\mathbf{A}^{\tau-1},A_{\tau})}{\sum_{y_{\tau}\in X_{\tau}}q(\mathbf{x}^{\tau-1},y_{\tau},\mathbf{A}^{\tau-1},X_{\tau})} \\
           & = q(x_1,A_1) \prod_{\tau=2}^T t^{\tau-1}_{\succ_{\tau}}\frac{q(\mathbf{x}^{\tau-1},x_{\tau},\mathbf{A}^{\tau-1},A_{\tau})}{q(\mathbf{x}^{\tau-1},\mathbf{A}^{\tau-1})} \\
           & = q(\mathbf{x}^T,\mathbf{A}^T)
        \end{split}
    \end{equation}
    Above, the first equality holds by the construction in our algorithm. The second equality holds by our second observation at the beginning of this proof. The third equality holds by canceling out common terms. The second case is when $q(\mathbf{x}^T,\mathbf{A}^T)=0$. In this case, there is some minimal length history $(\mathbf{x}^\mathcal{T},\mathbf{A}^\mathcal{T})$ contained by the history $(\mathbf{x}^T,\mathbf{A}^T)$ such that $q(\mathbf{x}^\mathcal{T},\mathbf{A}^\mathcal{T})=0$.
        \begin{equation}\label{verification0}
        \begin{split}
            & \sum_{\succ_1 \in I(x_1,A_1)} \cdots \sum_{\succ_T \in I(x_T,A_T)} \nu(\succ) \prod_{\tau=2}^T t^{\tau-1}_{\succ_{\tau}}(\mathbf{x}^{\tau-1},\succ^{\tau-1}) \\
           & = q(x_1,A_1) \prod_{\tau=2}^\mathcal{T} t^{\tau-1}_{\succ_{\tau}}\frac{q(\mathbf{x}^{\tau-1},x_{\tau},\mathbf{A}^{\tau-1},A_{\tau})}{\sum_{y_{\tau}\in X_{\tau}}q(\mathbf{x}^{\tau-1},y_{\tau},\mathbf{A}^{\tau-1},X_{\tau})}\prod_{\tau=\mathcal{T}+1}^T t^{\tau-1}_{\succ_{\tau}}(\mathbf{x}^{\tau-1},\succ^{\tau-1}) \\
           & = q(x_1,A_1) \prod_{\tau=2}^\mathcal{T} t^{\tau-1}_{\succ_{\tau}}\frac{q(\mathbf{x}^{\tau-1},x_{\tau},\mathbf{A}^{\tau-1},A_{\tau})}{q(\mathbf{x}^{\tau-1},\mathbf{A}^{\tau-1})}\prod_{\tau=\mathcal{T}+1}^T t^{\tau-1}_{\succ_{\tau}}(\mathbf{x}^{\tau-1},\succ^{\tau-1})  \\
           & = q(\mathbf{x}^\mathcal{T},\mathbf{A}^\mathcal{T}) \prod_{\tau=\mathcal{T}+1}^T t^{\tau-1}_{\succ_{\tau}}(\mathbf{x}^{\tau-1},\succ^{\tau-1}) \\
           & =0 = q(\mathbf{x}^T,\mathbf{A}^T)
        \end{split}
    \end{equation}
    The first three equalities hold for the same reason as in the positive case. The fourth equality holds due to $q(\mathbf{x}^\mathcal{T},\mathbf{A}^\mathcal{T})=0$. By Theorem \ref{uniquenessfullthm}, we know that our construction is a consumption dependent random utility representation of random joint choice rule $p$, and so we are done.
    
\end{proof}

\subsection{Proof of Theorem \ref{thm:dynamicsophistication}}
 We begin with a reminder that we only prove this result in the case of two periods even though the proof technique extends to more periods.
\begin{proof}
    To begin, $p$ can be represented by a distribution $\mu$ over preferences on two period consumption streams. Since $p$, satisfies marginality, $p(x,A)$ is well defined for first period choices. Fix a $y \in X$. Since $p(x,A)=\sum_{y \in \{y\}} p(x,y,A,\{y\})$, it then follows that $p(x,A)$ is governed by the induced distribution over rankings of alternatives $(x,y)$ for $x \in X$. To specify, restricting to sets of the form $A \times \{y\}$, our distribution over alternatives in $X \times X$ now only matters up to the ranking of alternatives $(x,y)$ for $x \in X$. Thus $\mu$ now induces some distribution over rankings of $(x,y)$ for $y$ fixed. We may vary $y$ and this distribution may differ, but it must be observationally equivalent to the distribution over rankings of $(x,y)$ as marginality is satisfied. As such, let $\nu(\succ)$ denote a distribution over preferences over $X$ which is given by the $\mu$'s induced distribution over $(x,y)$ for a chosen fixed $y$.

    Now we define some additional objects. Given a preference $\rhd$, let $\rhd_x$ denote the induced preference over second period alternatives given that the first period choice is $x$. Let $\mu(\{\rhd|y \rhd_x B\}| \rhd \text{ s.t. } (x,\cdot) \rhd A \times B)$ denote the probability of a preference $\rhd$ which chooses $y$ in from $B$ in the second period conditional on the event that $x$ is chosen from $A$ in the first period and $B$ is the second period set. If there are no positive probability events in this conditioning set, we set our conditional probability to zero. We now introduce a sequence of equalities which we will justify afterwards.
    \begin{equation}\label{DRUMCDRUMEquations}
        \begin{split}
            p(x,y,A,B) & = \mu(\{\rhd|(x,y) \rhd A \times B\}) \\
            & = \mu(\{\rhd | (x,\cdot) \rhd A \times B\}) \mu(\{\rhd|y \rhd_x B\}| \rhd \text{ s.t. } (x,\cdot) \rhd A \times B) \\
            & = \mu(\{\rhd | (x,\cdot) \rhd A \times B\}) \\
            & \times \sum_{B \subseteq B'} \mu(\{\rhd|X \setminus B' \rhd_x y \rhd_x B'\}| \rhd \text{ s.t. } (x,\cdot) \rhd A \times B) \\
            & = \nu(\{\succ| x \succ A\}) \sum_{B \subseteq B'} \mu(\{\rhd|X \setminus B' \rhd_x y \rhd_x B'\}| \rhd \text{ s.t. } (x,\cdot) \rhd A \times B) \\
            & = \sum_{A \subseteq A'} \nu(\{\succ| X \setminus A' \succ x \succ A'\}) \\
            & \times \sum_{B \subseteq B'} \mu(\{\rhd|X \setminus B' \rhd_x y \rhd_x B'\}| \rhd \text{ s.t. } (x,\cdot) \rhd A \times B) \\
            & = \sum_{A \subseteq A'} \sum_{B \subseteq B'} [\nu(\{\succ| X \setminus A' \succ x \succ A'\}) \\
            & \times \mu(\{\rhd|X \setminus B' \rhd_x y \rhd_x B'\}| \rhd \text{ s.t. } (x,\cdot) \rhd A \times B) ]
        \end{split}
    \end{equation}
    We now justify each equality. The first equality is the definition of a dynamic random utility representation. The second equality consists of breaking down a joint probability into an unconditional probability times a conditional probability. The third equality comes from the following observation. Note that every preference in the set $\{\rhd|X \setminus B' \rhd_x y \rhd_x B'\}$ and the set $\{\rhd|(x,\cdot) \rhd A \times B\}$ chooses $(x,y)$ from $A \times B'$. Since $(x,y)$ is still available in $A \times B$, these set of preferences also choose $(x,y)$ from $A \times B$. The fourth equality follows from our arguments in the first paragraph. The fifth equality is a standard combinatorial identity and holds for reasons similar to the third equality. The last equality simply shuffles the location of a summation term. Now recall our definition of M\"{o}bius inverse.
    \begin{equation*}
        p(x,y,A,B) = \sum_{A \subseteq A'} \sum_{B \subseteq B'} q(x,y,A',B')
    \end{equation*}
    After applying a simple induction argument using the last equality of Equation \ref{DRUMCDRUMEquations}, we get the following.
    \begin{equation}
        q(x,y,A',B') = \nu(\{\succ| X \setminus A' \succ x \succ A'\}) \mu(\{\rhd|X \setminus B' \rhd_x y \rhd_x B'\}| \rhd \text{ s.t. } (x,\cdot) \rhd A \times B)
    \end{equation}
    Since both $\nu$ and $\mu$ are probability distributions, we get that $q(x,y,A,B) \geq 0$. Since we assumed marginality holds at the onset, this means that both marginality and complete monotonicity hold, thus giving us consistency with CDRUM.
\end{proof}

\subsection{Proof of Theorem \ref{SICDRUMthm}}

Before proving Theorem \ref{SICDRUMthm}, we first extend the model, characterizing axiom, and result to the extended model considered in Appendix \ref{extendedmodel}. We then prove the extended result and take Theorem \ref{SICDRUMthm} as a special case of this extended result.

\begin{definition}
    A random joint choice rule $p$ is \textbf{consistent} with state independent consumption dependent random utility if there exists a probability distribution over preferences $\nu$ and a series of state independent transition functions $\{t^1,\dots,t^{T-1}\}$ such that the following holds.
    \begin{equation}\label{sicdrumfulldef}
        p(\mathbf{x}^T,\mathbf{A}^T)=\sum_{\succ_1 \in N(x_1,A_1)} \cdots \sum_{\succ_T \in N(x_T,A_T)} \nu(\succ_1) \prod_{\tau=2}^T t^{\tau-1}_{\succ_{\tau}}(\mathbf{x}^{\tau-1}) 
    \end{equation}
\end{definition}

\begin{axiom}[Choice Set Independence]\label{ChoiceSetIndFull}
    A random joint choice rule $p$ satisfies \textbf{choice set independence} if for each $\tau \in \{1, \dots, T-1\}$, for each $y \in B \subseteq X$, for each $(\mathbf{x}^\tau,\mathbf{A}^\tau)$ and $(\mathbf{x}^\tau,\mathbf{A'}^\tau)$ with $p(\mathbf{x}^\tau,\mathbf{A}^\tau)>0$ and $p(\mathbf{x}^\tau,\mathbf{A'}^\tau)>0$ we have $\frac{p(\mathbf{x}^\tau,y_{\tau+1},\mathbf{A}^\tau,B_{\tau+1})}{p(\mathbf{x}^\tau,\mathbf{A}^\tau)}=\frac{p(\mathbf{x}^\tau,y_{\tau+1},\mathbf{A'}^\tau,B_{\tau+1})}{p(\mathbf{x}^\tau,\mathbf{A'}^\tau)}$.
\end{axiom}

\begin{theorem}\label{SICDRUMFULL}
    A random joint choice rule $p$ is consistent with state independent consumption dependent random utility if and only if it satisfies Axioms \ref{marginalitygen}-\ref{ChoiceSetIndFull}.
\end{theorem}

Before proceeding with our proof of Theorem \ref{SICDRUMFULL}, we need one preliminary lemma.

\begin{lemma}\label{strongChoiceSetInd}
    Suppose that $p$ satisfies Axioms \ref{marginalitygen}-\ref{ChoiceSetIndFull}, then it is the case that for each $\tau \in \{1, \dots, T-1\}$, for each $y \in B \subseteq X$, for each $(\mathbf{x}^\tau,\mathbf{A}^\tau)$ and $(\mathbf{x}^\tau,\mathbf{A'}^\tau)$ with $q(\mathbf{x}^\tau,\mathbf{A}^\tau)>0$ and $q(\mathbf{x}^\tau,\mathbf{A'}^\tau)>0$ we have $\frac{q(\mathbf{x}^\tau,y_{\tau+1},\mathbf{A}^\tau,B_{\tau+1})}{q(\mathbf{x}^\tau,\mathbf{A}^\tau)}=\frac{q(\mathbf{x}^\tau,y_{\tau+1},\mathbf{A'}^\tau,B_{\tau+1})}{q(\mathbf{x}^\tau,\mathbf{A'}^\tau)}$.
\end{lemma}

\begin{proof}
    To begin, fix a $\tau \in \{1, \dots, T-1\}$, a $y \in B \subseteq X$, a $(\mathbf{x}^\tau,\mathbf{A}^\tau)$ and a $(\mathbf{x}^\tau,\mathbf{C}^\tau)$ with $q(\mathbf{x}^\tau,\mathbf{A}^\tau)>0$ and $q(\mathbf{x}^\tau,\mathbf{C}^\tau)>0$. Since $X$ is finite, we can choose $\mathbf{C}^\tau$ such that there exists no other $\mathbf{D}^\tau$ with $\mathbf{C}^\tau \subsetneq \mathbf{D}^\tau \subseteq \mathbf{X}^\tau$ and $q(\mathbf{x}^\tau, \mathbf{D}^\tau)>0$. Observe the following.
    \begin{equation}
        \begin{split}
            & q(\mathbf{x}^\tau,y_{\tau+1},\mathbf{A}^\tau,B_{\tau+1}) \\
            & = \sum_{A_1 \subseteq A'_1}\cdots \sum_{A_\tau \subseteq A'_\tau}\sum_{B \subseteq B'}\left(\prod_{i=1}^\tau(-1)^{|A'_i\setminus A_i|}\right)(-1)^{|B'\setminus B|}p(\mathbf{x}^\tau,y_{\tau+1},\mathbf{A'}^\tau,B'_{\tau+1}) \\
            &=\sum_{A_1 \subseteq A'_1}\cdots \sum_{A_\tau \subseteq A'_\tau}\sum_{B \subseteq B'}\left(\prod_{i=1}^\tau(-1)^{|A'_i\setminus A_i|}\right)(-1)^{|B'\setminus B|}p(\mathbf{x}^\tau,y_{\tau+1},\mathbf{C}^\tau,B'_{\tau+1})\frac{p(\mathbf{x}^\tau,\mathbf{A'}^\tau)}{p(\mathbf{x}^\tau,\mathbf{C}^\tau)} \\
            &=\sum_{A_1 \subseteq A'_1}\cdots \sum_{A_\tau \subseteq A'_\tau} \left(\prod_{i=1}^\tau(-1)^{|A'_i\setminus A_i|}\right)\frac{p(\mathbf{x}^\tau,\mathbf{A'}^\tau)}{p(\mathbf{x}^\tau,\mathbf{C}^\tau)} \sum_{B \subseteq B'}(-1)^{|B'\setminus B|}p(\mathbf{x}^\tau,y_{\tau+1},\mathbf{C}^\tau,B'_{\tau+1})\\
            &=\frac{q(\mathbf{x}^\tau,\mathbf{A}^\tau)}{q(\mathbf{x}^\tau,\mathbf{C}^\tau)}q(\mathbf{x}^\tau,y_{\tau+1},\mathbf{C}^\tau,B_{\tau+1})
        \end{split}
    \end{equation}
    Above, the first equality is due to the definition of M\"{o}bius inversion, the fact that set inclusion forms the Boolean algebra, and Proposition 5 and Corollary (Principle of Inclusion-Exclusion) from \citet{rota1964foundations}. The second equality holds due to choice set independence. The third equality is just a rearrangement. The fourth equality again holds due to the definition of M\"{o}bius inversion and the following argument. As there are no other $\mathbf{D}^\tau$ with $\mathbf{C}^\tau \subsetneq \mathbf{D}^\tau \subseteq \mathbf{X}^\tau$ and $q(\mathbf{x}^\tau, \mathbf{D}^\tau)>0$ and as complete monotonicity holds, it is the case that $p(\mathbf{x}^\tau,y_{\tau+1},\mathbf{D}^\tau,B'_{\tau+1})=0$ for all $\mathbf{C}^\tau \subsetneq \mathbf{D}^\tau$ and for all $B_{\tau+1} \subseteq B'_{\tau+1}$. While the typical M\"{o}bius inversion formula requires us to sum over all supersets of $B_{\tau+1}$ and $\mathbf{C}^\tau$, by the observation of the previous sentence, all terms with supersets of $\mathbf{C}^\tau$ drop out and our fourth equality holds. Thus the equality from Lemma \ref{strongChoiceSetInd} holds and we are done.
\end{proof}

We now proceed with our proof of Theorem \ref{SICDRUMFULL}.

\begin{proof}
    As necessity of Axioms \ref{marginalitygen} and \ref{compmongen} is shown in the proof of Theorem \ref{cdrumfull}, to show necessity here, all we need to do is show necessity of choice set independence. By the definition of a state independent consumption dependent random utility representation, we have the following.
    \begin{equation}
        \begin{split}
        & \frac{p(\mathbf{x}^\tau,y_{\tau+1},\mathbf{A}^\tau,B_{\tau+1})}{p(\mathbf{x}^\tau,\mathbf{A}^\tau)} \\
        & =\frac{\sum_{\succ_1 \in N(x_1,A_1)} \cdots \sum_{\succ_\tau \in N(x_\tau,A_\tau)}\sum_{\succ_{\tau+1} \in N(y_{\tau+1},B_{\tau+1})} \nu(\succ_1) t^{\tau}_{\succ_{\tau+1}}(\mathbf{x}^{\tau}) \prod_{\tau=2}^T t^{\tau-1}_{\succ_{\tau}}(\mathbf{x}^{\tau-1}) }{\sum_{\succ_1 \in N(x_1,A_1)} \cdots \sum_{\succ_T \in N(x_T,A_T)} \nu(\succ_1) \prod_{\tau=2}^T t^{\tau-1}_{\succ_{\tau}}(\mathbf{x}^{\tau-1}) } \\
        &=\sum_{\succ_{\tau+1} \in N(y_{\tau+1},B_{\tau+1})}t^{\tau}_{\succ_{\tau+1}}(\mathbf{x}^{\tau})
        \end{split}
    \end{equation}
    The second equality holds by canceling like terms. Note the the sum we are left with has no dependence on the prior periods' choice sets or preferences. Only the consumption terms enter into the the sum. Thus choice set independence holds.

    We now show sufficiency. As complete monotonicity and marginality hold, we know that our random joint choice rule has a consumption dependent random utility representation. All that is left to show is that the addition of choice set independence allows us to make this representation state independent. As complete monotonicity, marginality, and choice set independence hold, we know that the condition from Lemma \ref{strongChoiceSetInd} holds. In step 2 of Algorithm \ref{decompalg} where we normalize the total outflow from $X$ to be equal to one, our denominator/normalizing constant is exactly equal to the denominator in the expression from Lemma \ref{strongChoiceSetInd}. This means that, for every pair of graphs associated with histories $(\mathbf{x}^\tau,\mathbf{A}^\tau)$ and $(\mathbf{x}^\tau,\mathbf{B}^\tau)$ such that $q(\mathbf{x}^\tau,\mathbf{A}^\tau)>0$ and $q(\mathbf{x}^\tau,\mathbf{B}^\tau)>0$, the graphs that Algorithm \ref{decompalg} ends up decomposing are the same. As such, it immediately follows that the output $\nu$ in both of these cases are the same (or can be chosen to be the same). Thus, when we assign $t_\succ^\tau(\mathbf{x}^\tau,I(x_1,A_1),\dots,I(x_\tau,A_\tau))=\nu_{(\mathbf{x}^\tau,\mathbf{A}^\tau)}(\succ)$, as in the proof of Theorem \ref{cdrumfull}, $\nu$ is actually independent of $\mathbf{A}^\tau$, and $t$ is thus state independent. This shows that when complete monotonicity, marginality, and choice set independence hold, $p$ has a state independent consumption dependent random utility representation, and so we are done.
\end{proof}

\section{Proofs from Section \ref{sec:hypo}}

While we do not do so here, all of the results from Section \ref{sec:hypo} can be extended to the extended model introduced in Appendix \ref{extendedmodel} by asking that the analogues from Appendix \ref{app:axioms} of the equations discussed in Section \ref{sec:hypo} hold.

\subsection{Proof of Theorem \ref{HypoTestLinProThm}}
\begin{proof}
    
We begin by showing the equivalence between $(1)$ and $(2)$. Note that, by the definition of consumption dependent random utility, $(1)$ is equivalent to there existing some vector $r$ solving Equations \ref{ConicConsistency}-\ref{ConicPos} and $\sum_i r_i=1$. Now note that each row in matrix $E$ corresponds to a deterministic choice function and assigns total probability of one to choices from any choice set $A \times B$. Further, there are no negative entries in $E$. As our random joint choice rule $p$ also assigns total probability of one to choice from any choice set $A \times B$, any vector $r \geq 0$ which satisfies $rE=p$ must satisfy $\sum_i r_i=1$. Thus $\sum_i r_i=1$ is a redundant condition, and so we are done.

We now show the equivalence between $(1)$ and $(3)$. We begin with necessity. If our random joint choice rule $p$ is consistent with consumption dependent random utility, then there exists an extension of the random joint choice rule to $\mathcal{X}^2$ such that this extension satisfies marginality and complete monotonicity. Equation \ref{Qconsistincy} then holds as we are using $q$ to represent the M\"{o}bius inverse of our potentially full domain random joint choice rule. Recall that $q(x,y,X,X)=p(x,y,X,X)$. It immediately follows that Equation \ref{QInitialCon} holds. When marginality holds, recall that we define $q(x,A)=\sum_{y \in X}q(x,y,A,X)$. Further, when marginality holds, $p(x,A)$ is well-defined. As such, by Theorem 3.1 of \citet{turansick2023graphical}, Equation \ref{QinflowOutflow} holds. Equation \ref{Qpositivity} holds due to Theorem \ref{CDRUM}. Further, as marginality holds, by Lemma \ref{recursivity2p}, Equation \ref{Qrecursivity} holds.

We now move on to sufficiency. As Equation \ref{Qconsistincy} holds, our $q$ vector is equivalent to the M\"{o}bius inverse of some function $p$ which takes values on $\mathcal{X}^2$. As Equation \ref{Qrecursivity} holds, by Lemma \ref{recursivity2p}, we know that this $p$ function satisfies marginality (even if it is not a random joint choice rule). As such, $p(x,A)$ is well-defined. By Theorem 3.1 of \citet{turansick2023graphical}, by Equation \ref{QinflowOutflow} holding, and by $q(x,A)=\sum_{y\in X}q(x,y,A,X)$, $p(x,A)$ is a set-constant function as defined in \citet{turansick2023graphical}. By Equations \ref{Qpositivity} and \ref{QInitialCon} holding and by Corollaries 3.2 and 3.3 of \citet{turansick2023graphical}, it follows that $p(x,A)$ is in fact a random choice rule. By Equation \ref{Qpositivity}, each $p(x,y,A,B)$ is non-negative. As $\sum_{y \in B}p(x,y,A,B)=p(x,A)$ and as $p(x,A)$ defines a random choice rule, $\sum_{x\in A}\sum_{y\in B} p(x,y,A,B)=1$. This then means that $p(x,y,A,B)$ defines a full domain random joint choice rule. As Equation \ref{Qpositivity} encodes total monotonicity and Equation \ref{Qrecursivity} is equivalent to marginality holding, this random joint choice rule $p$ is consistent with consumption dependent random utility by Theorem \ref{CDRUM}, and so we are done.
\end{proof}

\subsection{Proof of Theorem \ref{FasterHypoTest}}

\begin{proof}
    As the set of constraints imposed by Equations \ref{Qconsistincy}-\ref{QInitialCon} contains the set of constraints imposed by Equations \ref{Qconsistincy}-\ref{Qpositivity} and \ref{QinflowOutflowLim}-\ref{QInitialConLim}, it immediately follows that there exists a solution to Equations \ref{Qconsistincy}-\ref{Qpositivity} and \ref{QinflowOutflowLim}-\ref{QInitialConLim} if there exists a solution to Equations \ref{Qconsistincy}-\ref{QInitialCon}. As such, we now show the other direction.

    As Equation \ref{Qconsistincy} holds, $q$ describes the M\"{o}bius inverse of a function $p$ which is defined for all of $\mathcal{X}^2$ and is consistent with our observed choice probabilities. Further, as Equation \ref{Qrecursivity} holds, we know that $p(x,A)$ is well-defined for this full domain function. We now proceed by induction. Our base case is when $A = X$. There are two cases. suppose that there is no set $X \times B \in \mathcal{X}_{lim}$. Then we ask Equation \ref{QInitialConLim} to hold and we know that $\sum_{x \in X} \sum_{y\in X}q(x,y,X,X)=\sum_{x\in X}\sum_{y\in X}p(x,y,X,X)=\sum_{x\in X}p(x,X)=1$. Further, by Equation \ref{Qpositivity} holding, we know that $p(x,X) \geq 0$. Our second case is when there exists some set $X \times B \in \mathcal{X}_{lim}$. Then we observe $p(x,y,X,B)$ and that $\sum_{y \in B}p(x,y,X,B)=1$. By Equation \ref{Qrecursivity}, it then follows that $\sum_{y \in X}p(x,y,X,X)=1$. Further, by Equation \ref{Qpositivity} holding, we know that $p(x,X) \geq 0$.

    We now move on to our induction argument. Suppose that for all $A'$ such that $A \subsetneq A'$ we know that $p(x,A')\geq 0$, $\sum_{x \in A'}p(x,A')=1$, and that $\sum_{A' \subseteq A''}q(x,A'')=p(x,A')$. It then follows that on the domain $\{A'|A \subsetneq A'\}$, $q$ is the M\"{o}bius inverse of $p$. There are again two cases. The first case is that there is some set $A \times B \in \mathcal{X}_{lim}$. Then, by Equation \ref{Qrecursivity}, we can define $p(x,A)=\sum_{y \in X}p(x,y,A,X)=\sum_{y \in B}p(x,y,A,B)=1$. By Equation \ref{Qconsistincy}, it then follows that $q$ is the M\"{o}bius inverse of $p$ at $A$ as well. Since choice is observed, we know that $p(x,y,A,B) \geq 0$ and thus $p(x,A)\geq 0$. We now move on to the second case. Suppose there is no set $A \times B \in \mathcal{X}_{lim}$. It then follows that we can choose $p(x,A)$ such that $\sum_{A \subseteq A'} q(x,A')=p(x,A)$. As we have thus far defined $p(x,A')=\sum_{y \in X}p(x,y,A',X)$, we know that $q(x,A)=\sum_{y \in X}q(x,y,A,X)$. It then follows from Equation \ref{QinflowOutflowLim}, Theorem 3.1 from \citet{turansick2023graphical}, and the fact that $\sum_{x\in A'}p(x,A')=1$ on $\{A'|A \subsetneq A'\}$ that $\sum_{x \in A}p(x,A)=1$. Further, by Equation \ref{Qpositivity}, we know that $p(x,A) \geq 0$.

    Thus far we have shown that when Equations \ref{Qconsistincy}-\ref{Qpositivity} and \ref{QinflowOutflowLim}-\ref{QInitialConLim} hold, there exists a random choice rule $p(x,A)$ (which happens to be consistent with classic random utility) that is consistent with our observed choice probabilities. As Equations \ref{Qrecursivity} and \ref{Qpositivity} hold, if there exists an extension to a full random joint choice rule $p$ consistent with our observed choice probabilities, then it is consistent with consumption dependent random utility. As such, all there is left to show is that we can extend our random joint choice rule on $\mathcal{X}_{lim}^2$ to a random joint choice rule on $\mathcal{X}^2$. When $p(x,y,A,B)$ is unobserved, we can choose it so that $\sum_{A \subseteq A'}\sum_{B \subseteq B'}q(x,y,A',B')=p(x,y,A,B)$. When $p(x,y,A,B)$ is observed, Equation \ref{Qconsistincy} holds. Thus, we know that $q$ is the M\"{o}bius inverse of the full domain $p$. Thus far in our construction, we have defined $p(x,A)$ using $\sum_{y \in X}p(x,y,A,X)$ and have thus made sure that $\sum_{A\subseteq A'}q(x,y,A',X)=p(x,y,A,X)$. By Equation \ref{Qrecursivity} and Lemma \ref{recursivity2p}, it follows that $\sum_{y \in B}p(x,y,A,B)=\sum_{y \in X}p(x,y,A,X)=p(x,A)$. As $\sum_{x\in A}p(x,A)=1$, it then follows that $\sum_{x\in A}\sum_{y \in B}p(x,y,A,B)=1$. By Equation \ref{Qpositivity}, it follows that $p(x,y,A,B) \geq 0$. Thus any $q$ which is a solution to Equations \ref{Qconsistincy}-\ref{Qpositivity} and \ref{QinflowOutflowLim}-\ref{QInitialConLim} defines a full domain random joint choice rule which is consistent with consumption dependent random utility, and so we are done.
\end{proof}

\subsection{Proof of Theorem \ref{dimThm}}
Our proof of Theorem \ref{dimThm} depends on an different graphical representation than the one proposed in Appendix \ref{extendedmodel}. It takes the graph corresponding to the null history as its base and then adds the graph corresponding to the history $(x,A)$ at the edge in null history graph connecting $A$ to $A \setminus \{x\}$. Formally, our composite graph has a collection of nodes. These nodes are of two types. The first type is indexed by $2^X$. The second set are indexed by $\{(x,A,B)|x \in A \in 2^X,B \in 2^X\}$. The edge set is formed as follows.
\begin{itemize}
    \item For a node $A \neq \emptyset$, there is an edge from $A$ to $(x,C,B)$ if $C =A$ and $B = X$.
    \item For the node $\emptyset$, there is an edge from $\emptyset$ to $X$.
    \item For a node $(x,A,B)$ with $B \neq \emptyset$, there is an edge from $(x,A,B)$ to $(x,A,C)$ if $C = B \setminus \{y\}$ for some $y \in B$.
    \item For a node $(x,A,\emptyset)$, there is an edge from $(x,A,\emptyset)$ to $A \setminus \{x\}$.
\end{itemize}
We now populate each edge with an edge capacity.
\begin{itemize}
    \item For an edge from $A$ to $(x,A,X)$, assign $q(x,A)$ as edge capacity.\footnote{Recall that $q(x,A)$ is the M\"{o}bius inverse of $p(x,A)$ which is the first period random choice rule when our rjcr satisfies marginality.}
    \item For the edge from $\emptyset$ to $X$, assign $1$ as edge capacity.
    \item For the edge from $(x,A,B)$ to $(x,A,B\setminus \{y\})$, assign $q(x,y,A,B)$ as edge capacity.
    \item For the edge from $(x,A,\emptyset)$ to $A \setminus \{x\}$, assign $q(x,A)$ as the edge capacity.
\end{itemize}

In Section \ref{sec:hypo}, we introduced the extreme points of CDRUM as tuples of preferences, $(\succ,\succ_{x_1},\dots,\succ_{x_{|X|}})$, one dictating choice in the first period and the others dictating choice in the second period conditional on first period choice. We now note that there is a bijection between the set of these preference tuples and paths from $X$ to $\emptyset$ in our composite graph. Ignoring nodes of the form $(x,A,B)$ for now, a path from $X$ to $\emptyset$ corresponds to a preference just in the way it did in Appendix \ref{extendedmodel}. This is the preference that governs first period choice. Now we focus on the nodes of the form $(x,A,B)$ for a fixed $(x,A)$. As before, a path from $(x,A,X)$ to $(x,A,\emptyset)$ is bijectively associated with a preference. This is the preference that governs second period choice conditional on choosing $x$ in the first period. In total, there is a bijection between $X$ to $\emptyset$ paths in our composite graph and a tuples of preferences of the form $(\succ,\succ_{x_1},\dots,\succ_{x_{|X|}})$. Data induced by choice according to $(\succ,\succ_{x_1},\dots,\succ_{x_{|X|}})$ leaves $1$ as the edge capacity on each edge in the path bijectively associated with $(\succ,\succ_{x_1},\dots,\succ_{x_{|X|}})$ and $0$ on all other edges. We now proceed with our proof which extends the techniques used in \citet{CHAMBERS2025}.

\begin{proof}
    Begin by recalling that the dimension of a vector space is the maximum size of a set of linearly independent vectors in that vector space. Now we can think of a vector $p$ which encodes a rjcr. Consider a linearly independent set $\{p_i\}$. Since M\"{o}bius inversion is a bijective linear transformation, $\{p_i\}$ is linearly independent if and only if $\{q_i\}$, the set of vectors of the M\"{o}bius inverses of $\{p_i\}$, is linear independent. By Proposition 3.2 of \citet{CHAMBERS2025}, $\{q_i\}$ is linearly independent if and only if $\{(q_i,1)\}$ is linearly independent. Each $(q_i,1)$ corresponds to the flow representation of some rjcr on our composite graph. Terminology in the rest of this proof is as in \citet{berge2001theory}.

    By Theorem 3 of Chapter 4 of \citet{berge2001theory}, the cyclomatic number of the composite graph is equal to the dimension of the space spanned by indicator functions of circuits of the composite graph. Now, the indicator function of any circuit of this diagram must be a linear combination of indicator functions of circuits corresponding to preference tuples.  The reasoning is straightforward:  by construction, every circuit must pass through the edge connecting $\emptyset$ to $X$.  A circuit which passes through this edge only once, as discussed earlier, corresponds to a specific tuple of preferences, $(\succ,\succ_{x_1},\dots,\succ_{x_{|X|}})$.  A circuit passing through it $k$ times corresponds to a concatenation of $k$ circuits passing through this edge only once.  Hence, the indicator function of the circuit passing through this edge $k$ times is the sum of the indicator functions of the $k$ circuits which pass through the edge only once, each of which correspond to a tuple of preferences $(\succ,\succ_{x_1},\dots,\succ_{x_{|X|}})$. Hence, the dimension of the circuit space of the composite graph is the same as the dimension of the space spanned by extreme points of CDRUM.

The cyclomatic number of a strongly connected graph, as our composite graph is, according to \citet{berge2001theory}, is defined by $E-N+1$, where $E$ is the number of edges and $N$ is the number of nodes. There are $2^{|X|}$ nodes of the form $A$ in our composite graph and $|X|(2^{|X|-1})2^{|X|}$ nodes of the form $(x,A,B)$ in our composite graph. To see the latter statement, observe that we have $2^{|X|}$ nodes of the form $(x,A,B)$ for a fixed $(x,A)$ and $|X|(2^{|X|-1})$ unique $(x,A)$ with $x \in A$ (see the proof of Theorem 3.1 in \citet{CHAMBERS2025}). There are $|X|(2^{|X|-1})$ edges going from nodes of the form $A$ to nodes of the form $(x,A,X)$ and another $|X|(2^{|X|-1})$ going from nodes of the form $(x,A,\emptyset)$ to nodes of the form $A \setminus \{x\}$. Once again, there are $|X|(2^{|X|-1})$ unique $(x,A)$ with $x \in A$, and for each of these there are $|X|(2^{|X|-1})$ edges going from nodes of the form $(x,A,B)$ to nodes of the form $(x,A,B \setminus \{y\})$, giving us a total of $(|X|(2^{|X|-1}))^2$ edges of this form. Finally, there is an edge from $\emptyset$ to $X$ giving us one additional edge. Taking these node and edge totals and applying them to the cyclomatic number equation gives us exactly $(|X|2^{|X|-1})^2-|X|2^{2|X|-1}+(|X|-1)2^{|X|}+2$. As stated earlier, the cyclomatic number of our graph is the dimension of the space spanned by indicator functions of circuits of the composite graph. As argued in the previous paragraph, this space is the same as the space spanned by the indicator functions of minimal circuits of the composite graph. However, the indicator function of a minimal circuit of the composite graphs is exactly the vector $(q,1)$ induced by choice by some extreme point of CDRUM, and so we are done.
\end{proof}

\bibliographystyle{ecta}
\bibliography{mrum}

\begin{thebibliography}{43}
\newcommand{\enquote}[1]{``#1''}
\expandafter\ifx\csname natexlab\endcsname\relax\def\natexlab#1{#1}\fi

\bibitem[\protect\citeauthoryear{Apesteguia, Ballester, and Lu}{Apesteguia et~al.}{2017}]{apesteguia2017single}
\textsc{Apesteguia, J., M.~A. Ballester, and J.~Lu} (2017): \enquote{Single-Crossing Random Utility Models,} \emph{Econometrica}, 85, 661--674.

\bibitem[\protect\citeauthoryear{Becker and Murphy}{Becker and Murphy}{1988}]{becker1988theory}
\textsc{Becker, G.~S. and K.~M. Murphy} (1988): \enquote{A theory of rational addiction,} \emph{Journal of political Economy}, 96, 675--700.

\bibitem[\protect\citeauthoryear{Berg, Christensen, and Ressel}{Berg et~al.}{1984}]{berg1984harmonic}
\textsc{Berg, C., J.~P.~R. Christensen, and P.~Ressel} (1984): \emph{Harmonic analysis on semigroups: theory of positive definite and related functions}, vol. 100, Springer.

\bibitem[\protect\citeauthoryear{Berge}{Berge}{2001}]{berge2001theory}
\textsc{Berge, C.} (2001): \emph{The theory of graphs}, Courier Corporation.

\bibitem[\protect\citeauthoryear{Block and Marschak}{Block and Marschak}{1959}]{block1959random}
\textsc{Block, H.~D. and J.~Marschak} (1959): \enquote{Random Orderings and Stochastic Theories of Response,} Tech. rep., Cowles Foundation for Research in Economics, Yale University.

\bibitem[\protect\citeauthoryear{Bornstein and D'Agostino}{Bornstein and D'Agostino}{1992}]{bornstein1992stimulus}
\textsc{Bornstein, R.~F. and P.~R. D'Agostino} (1992): \enquote{Stimulus recognition and the mere exposure effect.} \emph{Journal of personality and social psychology}, 63, 545.

\bibitem[\protect\citeauthoryear{Carroll, Overland, and Weil}{Carroll et~al.}{2000}]{carroll2000saving}
\textsc{Carroll, C.~D., J.~Overland, and D.~N. Weil} (2000): \enquote{Saving and growth with habit formation,} \emph{American Economic Review}, 90, 341--355.

\bibitem[\protect\citeauthoryear{Chambers, Masatlioglu, and Turansick}{Chambers et~al.}{2024}]{chambers2024correlated}
\textsc{Chambers, C.~P., Y.~Masatlioglu, and C.~Turansick} (2024): \enquote{Correlated choice,} \emph{Theoretical Economics}, 19, 1087--1117.

\bibitem[\protect\citeauthoryear{Chambers and Turansick}{Chambers and Turansick}{2025}]{CHAMBERS2025}
\textsc{Chambers, C.~P. and C.~Turansick} (2025): \enquote{The Limits of Identification in Discrete Choice,} \emph{Games and Economic Behavior}.

\bibitem[\protect\citeauthoryear{Clark}{Clark}{1996}]{clark1996random}
\textsc{Clark, S.~A.} (1996): \enquote{The random utility model with an infinite choice space,} \emph{Economic Theory}, 7, 179--189.

\bibitem[\protect\citeauthoryear{Deb and Renou}{Deb and Renou}{2021}]{deb2021dynamic}
\textsc{Deb, R. and L.~Renou} (2021): \enquote{Dynamic Choices and Common Learning,} \emph{arXiv preprint arXiv:2105.03683}.

\bibitem[\protect\citeauthoryear{Demuynck and Potoms}{Demuynck and Potoms}{2022}]{demuynck2022testing}
\textsc{Demuynck, T. and T.~Potoms} (2022): \emph{Testing revealed preference models with unobserved randomness: a column generation approach}, ECARES.

\bibitem[\protect\citeauthoryear{Duraj}{Duraj}{2018}]{duraj2018dynamic}
\textsc{Duraj, J.} (2018): \enquote{Dynamic Random Subjective Expected Utility,} Unpublished.

\bibitem[\protect\citeauthoryear{Falmagne}{Falmagne}{1978}]{falmagne1978representation}
\textsc{Falmagne, J.-C.} (1978): \enquote{A Representation Theorem for Finite Random Scale Systems,} \emph{Journal of Mathematical Psychology}, 18, 52--72.

\bibitem[\protect\citeauthoryear{Fang, Santos, Shaikh, and Torgovitsky}{Fang et~al.}{2023}]{fang2023inference}
\textsc{Fang, Z., A.~Santos, A.~M. Shaikh, and A.~Torgovitsky} (2023): \enquote{Inference for Large-Scale Linear Systems With Known Coefficients,} \emph{Econometrica}, 91, 299--327.

\bibitem[\protect\citeauthoryear{Fiorini}{Fiorini}{2004}]{fiorini2004short}
\textsc{Fiorini, S.} (2004): \enquote{A Short Proof of a Theorem of Falmagne,} \emph{Journal of Mathematical Psychology}, 48, 80--82.

\bibitem[\protect\citeauthoryear{Frick, Iijima, and Strzalecki}{Frick et~al.}{2017}]{frick2017dynamic}
\textsc{Frick, M., R.~Iijima, and T.~Strzalecki} (2017): \enquote{Dynamic Random Utility,} \emph{Working Paper}.

\bibitem[\protect\citeauthoryear{Frick, Iijima, and Strzalecki}{Frick et~al.}{2019}]{frick2019dynamic}
---\hspace{-.1pt}---\hspace{-.1pt}--- (2019): \enquote{Dynamic Random Utility,} \emph{Econometrica}, 87, 1941--2002.

\bibitem[\protect\citeauthoryear{Fudenberg and Strzalecki}{Fudenberg and Strzalecki}{2015}]{fudenberg2015dynamic}
\textsc{Fudenberg, D. and T.~Strzalecki} (2015): \enquote{Dynamic Logit with Choice Aversion,} \emph{Econometrica}, 83, 651--691.

\bibitem[\protect\citeauthoryear{Gul and Pesendorfer}{Gul and Pesendorfer}{2001}]{gul2001temptation}
\textsc{Gul, F. and W.~Pesendorfer} (2001): \enquote{Temptation and self-control,} \emph{Econometrica}, 69, 1403--1435.

\bibitem[\protect\citeauthoryear{Gul and Pesendorfer}{Gul and Pesendorfer}{2006}]{gul2006random}
---\hspace{-.1pt}---\hspace{-.1pt}--- (2006): \enquote{Random expected utility,} \emph{Econometrica}, 74, 121--146.

\bibitem[\protect\citeauthoryear{Havranek, Rusnak, and Sokolova}{Havranek et~al.}{2017}]{havranek2017habit}
\textsc{Havranek, T., M.~Rusnak, and A.~Sokolova} (2017): \enquote{Habit formation in consumption: A meta-analysis,} \emph{European economic review}, 95, 142--167.

\bibitem[\protect\citeauthoryear{Heckman}{Heckman}{1978}]{heckman1978simple}
\textsc{Heckman, J.~J.} (1978): \enquote{Simple statistical models for discrete panel data developed and applied to test the hypothesis of true state dependence against the hypothesis of spurious state dependence,} in \emph{Annales de l'INSEE}, JSTOR, 227--269.

\bibitem[\protect\citeauthoryear{Heckman}{Heckman}{1981}]{heckman1981heterogeneity}
---\hspace{-.1pt}---\hspace{-.1pt}--- (1981): \enquote{Heterogeneity and state dependence,} in \emph{Studies in labor markets}, University of Chicago Press, 91--140.

\bibitem[\protect\citeauthoryear{Hoch and Deighton}{Hoch and Deighton}{1989}]{hoch1989managing}
\textsc{Hoch, S.~J. and J.~Deighton} (1989): \enquote{Managing what consumers learn from experience,} \emph{Journal of marketing}, 53, 1--20.

\bibitem[\protect\citeauthoryear{Hoch and Ha}{Hoch and Ha}{1986}]{hoch1986consumer}
\textsc{Hoch, S.~J. and Y.-W. Ha} (1986): \enquote{Consumer learning: Advertising and the ambiguity of product experience,} \emph{Journal of consumer research}, 13, 221--233.

\bibitem[\protect\citeauthoryear{Kashaev, Aguiar, Pl\'{a}vala, and Gauthier}{Kashaev et~al.}{2023}]{kashaev2022nonparametric}
\textsc{Kashaev, N., V.~H. Aguiar, M.~Pl\'{a}vala, and C.~Gauthier} (2023): \enquote{Dynamic and Stochastic Rational Behavior,} \emph{arXiv preprint arXiv:2302.04417}.

\bibitem[\protect\citeauthoryear{Kitamura and Stoye}{Kitamura and Stoye}{2018}]{kitamura2018nonparametric}
\textsc{Kitamura, Y. and J.~Stoye} (2018): \enquote{Nonparametric Analysis of Random Utility Models,} \emph{Econometrica}, 86, 1883--1909.

\bibitem[\protect\citeauthoryear{Kono, Saito, and Sandroni}{Kono et~al.}{2023}]{kono2023axiomatization}
\textsc{Kono, H., K.~Saito, and A.~Sandroni} (2023): \enquote{Axiomatization of Random Utility Model with Unobservable Alternatives,} \emph{arXiv preprint arXiv:2302.03913}.

\bibitem[\protect\citeauthoryear{Li}{Li}{2022}]{li2021axiomatization}
\textsc{Li, R.} (2022): \enquote{An Axiomatization of Stochastic Utility,} \emph{arXiv preprint arXiv:2102.00143}.

\bibitem[\protect\citeauthoryear{Lu and Saito}{Lu and Saito}{2018}]{lu2018random}
\textsc{Lu, J. and K.~Saito} (2018): \enquote{Random Intertemporal Choice,} \emph{Journal of Economic Theory}, 177, 780--815.

\bibitem[\protect\citeauthoryear{McFadden and Richter}{McFadden and Richter}{1990}]{mcfadden1990stochastic}
\textsc{McFadden, D. and M.~K. Richter} (1990): \enquote{Stochastic Rationality and Revealed Stochastic Preference,} \emph{Preferences, Uncertainty, and Optimality, Essays in Honor of Leo Hurwicz, Westview Press: Boulder, CO}, 161--186.

\bibitem[\protect\citeauthoryear{Pennesi}{Pennesi}{2021}]{pennesi2021intertemporal}
\textsc{Pennesi, D.} (2021): \enquote{Intertemporal Discrete Choice,} \emph{Journal of Economic Behavior \& Organization}, 186, 690--706.

\bibitem[\protect\citeauthoryear{Pollak}{Pollak}{1970}]{pollak1970habit}
\textsc{Pollak, R.~A.} (1970): \enquote{Habit formation and dynamic demand functions,} \emph{Journal of political Economy}, 78, 745--763.

\bibitem[\protect\citeauthoryear{Rota}{Rota}{1964}]{rota1964foundations}
\textsc{Rota, G.-C.} (1964): \enquote{On the Foundations of Combinatorial Theory I. Theory of M{\"o}bius Functions,} \emph{Zeitschrift f{\"u}r Wahrscheinlichkeitstheorie und verwandte Gebiete}, 2, 340--368.

\bibitem[\protect\citeauthoryear{Rust}{Rust}{1987}]{rust1987}
\textsc{Rust, J.} (1987): \enquote{Optimal Replacement of GMC Bus Engines: An Empirical Model of Harold Zurcher,} \emph{Econometrica}, 57, 999--1033.

\bibitem[\protect\citeauthoryear{Smeulders, Cherchye, and De~Rock}{Smeulders et~al.}{2021}]{smeulders2021nonparametric}
\textsc{Smeulders, B., L.~Cherchye, and B.~De~Rock} (2021): \enquote{Nonparametric Analysis of Random Utility Models: Computational Tools for Statistical Testing,} \emph{Econometrica}, 89, 437--455.

\bibitem[\protect\citeauthoryear{Strack and Taubinsky}{Strack and Taubinsky}{2021}]{strack2021dynamic}
\textsc{Strack, P. and D.~Taubinsky} (2021): \enquote{Dynamic Preference “Reversals” and Time Inconsistency,} Tech. rep., National Bureau of Economic Research.

\bibitem[\protect\citeauthoryear{Strzalecki}{Strzalecki}{2024}]{strzalecki2024stochastic}
\textsc{Strzalecki, T.} (2024): \emph{Stochastic choice theory}, Cambridge University Press.

\bibitem[\protect\citeauthoryear{Turansick}{Turansick}{2023}]{turansick2023graphical}
\textsc{Turansick, C.} (2023): \enquote{On Graphical Methods in Stochastic Choice,} \emph{arXiv preprint arXiv:2303.14249}.

\bibitem[\protect\citeauthoryear{Turansick}{Turansick}{2024}]{turansick2023alternative}
---\hspace{-.1pt}---\hspace{-.1pt}--- (2024): \enquote{An alternative approach for nonparametric analysis of random utility models,} \emph{arXiv preprint arXiv:2303.14249}.

\bibitem[\protect\citeauthoryear{Yerkes and Dodson}{Yerkes and Dodson}{1908}]{yerkes1908relation}
\textsc{Yerkes, R.~M. and J.~D. Dodson} (1908): \enquote{The Relation of Strength of Stimulus to Rapidity of Habit-formation,} \emph{Journal of Comparative and Neurological Psychology}, 18.

\bibitem[\protect\citeauthoryear{Zajonc}{Zajonc}{2001}]{zajonc2001mere}
\textsc{Zajonc, R.~B.} (2001): \enquote{Mere exposure: A gateway to the subliminal,} \emph{Current directions in psychological science}, 10, 224--228.

\end{thebibliography}

\end{document}